\documentclass[11pt]{article} 

\usepackage[utf8]{inputenc} 

\usepackage{geometry} 
\usepackage{graphicx} 
\usepackage{float}
\usepackage[export]{adjustbox}

\usepackage{subcaption}
\usepackage{multirow}
\usepackage{longtable}

\usepackage{gensymb}
\usepackage{amsmath,amssymb}

\usepackage{booktabs} 
\usepackage{array} 
\usepackage{paralist} 
\usepackage{verbatim} 
\usepackage{subfig} 

\usepackage{fancyhdr} 
\usepackage{sectsty}
\allsectionsfont{\sffamily\mdseries\upshape} 

\usepackage[nottoc,notlof,notlot]{tocbibind} 
\usepackage[titles,subfigure]{tocloft} 

\newcommand{\kBT}{$k_\textrm BT$}
\newcommand{\M}{\textsc{m}}

\newcommand{\etai}[1]{\eta_\textrm{#1}}

\newcommand{\DDM}{\textsc{ddm}}
\newcommand{\MQ}{\textsc{mq}}
\newcommand\ph{$\phantom{l}$} 

\title{\vspace{-1cm}Probing microsecond dynamics of disperse nanoparticles at oil-water interfaces using nanometer localization precision total-internal-reflection dark-field microscopy}
\author{A. Y. Schumacher}
\date{} 

\begin{document}
\maketitle
\begin{abstract}
Particles at fluid-fluid interfaces have a wide range of important applications in industry, technology and science. Questions remain about what governs adsorption dynamics and inter-particle interactions, and how mixtures -- both of different-sized particles, and particles and surfactants -- behave. I probe the dynamics of nanoparticles at oil-water interfaces using high-speed total-internal-reflection dark-field (\textsc{tir-df}) microscopy, comparing gold particles of diameters 20, 40 and 80 nm, and investigating the effect of an added surfactant. I built a \textsc{tir-df} microscope with a spatiotemporal resolution of up to 1~nm at 20~$\mu$s, and characterized the particle motion by direct imaging for the diffusive motion within the interfacial plane, and attenuation of light scattering and in-plane motion for out-of-plane motion. The addition of surfactant was found to qualitatively alter the interaction between the particles and the interface: Adsorptions became reversible and the size trend in diffusivity was reversed (larger particles diffused faster than smaller ones). The combination of this and the rarity of adsorption events of larger particles leads me to propose size-dependent barriers to adsorption at surfactant interfaces. In addition, at surfactant interfaces, many particles' diffusive behavior changed over the course of a trajectory, indicating several different immersion states that are stable on the experimental timescale ($\approx$1~s). This study shows that there remain qualitatively new dynamics of nanoparticles at oil-water interfaces to be explored. These dynamics can only be resolved at a high spatial resolution, and many of them only at high imaging speed, underlining the importance of novel imaging approaches probing the barriers of achievable spatiotemporal resolution.
\end{abstract}

\section{Introduction}

Micro- and nano-sized particles are known to self-assemble into structured layers at liquid-liquid interfaces. This behavior has been used to synthesize thin films as photonic materials or microelectronic device components that would otherwise be very difficult or impossible to make. \cite{Hu2012, films_mater} The thermodynamic driving force of particle adsorption to the interface is a reduction in interfacial tension. \cite{Pieranski, Lin2003} In addition to films, this effect can be used to stabilize emulsions (so-called Pickering emulsions), another class of materials that is highly relevant in industry. \cite{Binks2017, McGorty2010, bijels} Once at the interface, inter-particle forces are determined by a complex interplay of electrostatics and geometric considerations, as well as particle shape, concentration and the concentration of other additives such as salt. \cite{BresmeOettel2007, Stebe_curvature}\\

\begin{figure}[H]
\vspace{.3cm}
\begin{subfigure}[t]{.23\textwidth}
\caption{}\label{fig:intro_particle}
\raisebox{-0.83\height}[0pt][0pt]{\includegraphics[height=3.4cm]{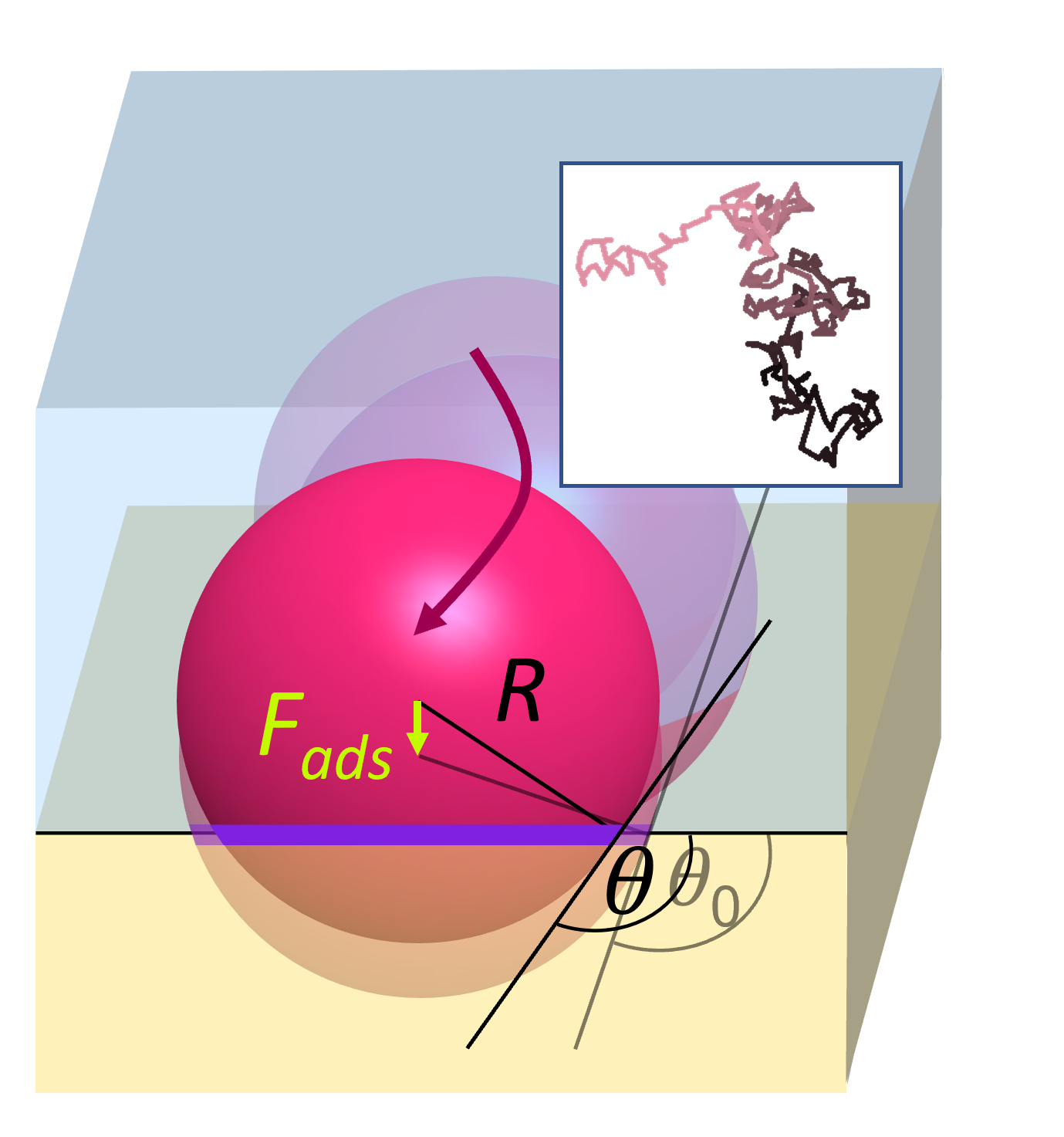}}
\end{subfigure}
\begin{subfigure}[t]{.38\textwidth}
\caption{}
\includegraphics[height=3.1cm]{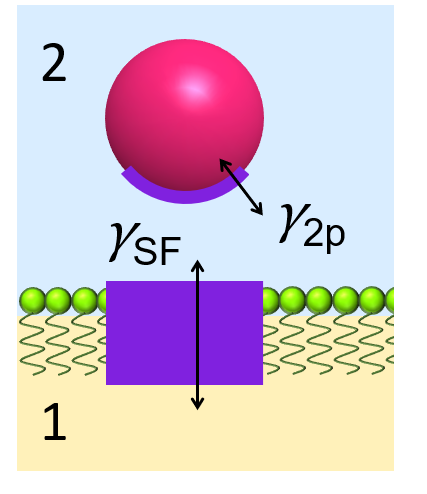}
\includegraphics[height=3.05cm,trim={0 0 0 0.1cm},clip]{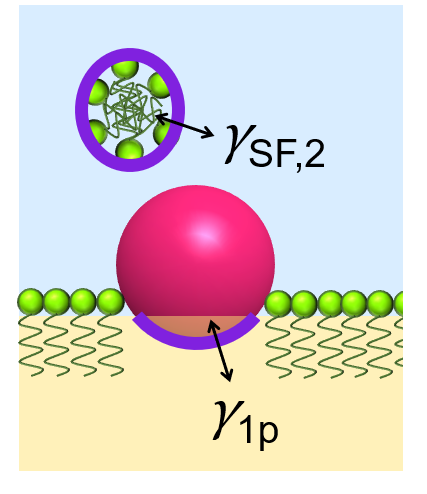}
\end{subfigure}
\begin{subfigure}[t]{.38\textwidth}
\caption{}
\includegraphics[height=3cm,trim={0 0 0.2cm 0.2cm},clip]{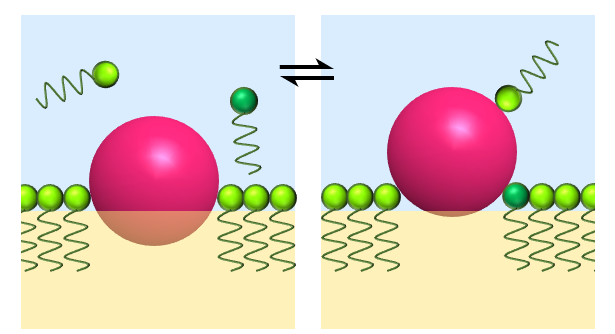}
\end{subfigure}
\caption{\textbf{Out-of-plane dynamics of particles at liquid-liquid interfaces and selected effects of surfactants on them.} a) Schematic trajectory of a particle at an oil-water interface, 2d projection in inset. Immersion angle and equilibrium immersion angle $\theta$ and $\theta_0$, and the force $F_\textrm{ads}$ resulting from a particle not being at the equilibrium immersion angle. b) Adsorption of a particle to a surfactant-laden interface (before and after). The adsorption energy is determined by a change in interfacial energy, with terms from changing interface areas are calculated via $E_\textrm i=\gamma_\textrm iA$. $\gamma_\textrm i$ is the interfacial tension and $A$ is the interfacial area. c) Addition of surfactant makes it thermodynamically and kinetically more favorable for the particle in case (a) to move away from its equilibrium adsorption angle, where the equilibrium angle is furthermore altered by this addition.}
\label{fig:intro}
\end{figure}

The adsorption energy calculated from changes in interfacial energies increases quadratically with particle size, \cite{Pieranski} and is often hundreds or even millions of times the thermal energy even for micron-sized particles. However, relaxation towards the equilibrium contact angle once they breach the interface can be slow, suggesting metastable states and hinting at the possibility of kinetic control. \cite{KazManoharan2011} For nanoparticles, the equilibrium adsorption energy is closer to the thermal energy, and for some combinations of particle size, temperature and interfacial energies between the three phases (two fluids and the particle), adsorbed particles can move perpendicular to the interface and even desorb \cite{JPCL} (see Fig.~\ref{fig:intro}). This effect enables size-selective particle assembly in 3D, with smaller particles being replaced by larger ones at the interface in a dynamic system. \cite{Lin2003}

Besides allowing for desorption and the increased control that comes with that, nanoparticles are of special interest because of their size. A significant fraction of their atoms are located at the surface, which qualitatively changes their material properties compared to larger-scale matter.~\cite{quantum_nanoscience} Observed effects include increased chemical reactivity (see for example refs.~\cite{battery_nanowires} and \cite{nanoCO2reduction}), tunable and intensified electromagnetic response (e.g. ref.~\cite{NatureNanoTao}) and improved (bio-)mechanical properties (e.g. refs.~\cite{windmills, bones}). The use of liquid-liquid interfaces is a potentially very simple technique for manufacturing these important materials, and this has become an important topic in materials science.~\cite{Shi2018} However, the same size-related effects that give nanoparticles interesting properties can complicate the description of their behavior: Classical theories assume non-polarizability and spherical shape of particles and negligible size of solvent molecules compared to the particle, and while these assumptions often hold for microparticles, they break down for nanoparticles.~\cite{nano_nonadditivity} The dynamics of nanoparticles are also more difficult to access experimentally because of nanoparticles' faster motion and lower scattering and fluorescence intensity, and hence are not as well characterized. Understanding the dynamics of nanoparticles at the interface would aid in controlling synthesis processes, giving access to new materials and the understanding of quantum and size-related effects that become visible at these scales.\\


In this work, I characterize the out-of-plane motion of disperse nanoparticles at oil-water interfaces in the absence and presence of an added surfactant, \textit{n}-dodecyl-$\beta$-\textsc{D}-maltoside (\DDM{}). In avoiding inter-particle interactions, I aim to gain insights into the interaction between particles and the interface. The surfactant is expected to enhance the stability of existing metastable immersion states and shift the desorption potential towards larger particle sizes \cite{Smits2019}, which are easier to image. Surfactants have also, however, been shown to alter particle wetting properties \cite{Binks2013} and lead to stronger particle adsorption via cooperative binding effects.~\cite{Whitby2012}\\

The most reliable way to study the dynamics of nanoparticles at interfaces is to observe them directly (single-particle tracking). A relatively non-invasive way to achieve this is to probe with electromagnetic radiation in the visible spectrum, i.e. with an optical microscope. Atomic force microscopy (AFM) and electron microscopy have a higher spatial resolution, but are not suitable because the liquid-liquid interface is inaccessible. Furthermore, the time resolution of AFM is too low. Typical methods for single-particle tracking include \textsc{tirf} microscopy \cite{SPTreview,Yildiz}, which yields high signal-to-noise ratios by suppressing background signals resulting from scattering, allowing small particles to be tracked accurately. However, fluorescence methods \cite{fastflu} are limited in localization precision and time resolution by bleaching and photoblinking. In addition, single quantum emitters have a limited photon emission rate due to the finite lifetime of the excited state. The length scale that is relevant for particle self-assembly is on the order of the particle diameter or smaller, which imposes a requirement for high localization precision, along with a high time resolution.\\

The characteristic time scale for diffusive motion is given by Einstein's diffusion equation $\langle \Delta x^2 \rangle = 4D\tau_\textrm d$, where $\Delta x$ is the particle displacement in the time interval $\tau_\textrm d$, the brackets denote an ensemble average. $D$ is the diffusion coefficient, and can be estimated using a modified Stokes-Einstein equation:

\begin{equation}
D=\frac{\textrm\kBT{}}{6\pi R_\textrm{h}}\frac{2}{\etai{}{1}+\etai{}{2}+(\etai{}{1}-\etai{}{2})\cos\theta},
\label{eq:StokesEinstein}
\end{equation}

where $k_\textrm B$ is the Boltzmann constant, $T$ is the temperature, $R_\textrm h$ is the hydrodynamic radius of the particle, the $\eta$ variables represent viscosities of the surrounding medium and $\theta$ is the contact angle as defined in Fig.~\ref{fig:intro_particle}. For particles with diameters of 10 and 100~nm, using diffusion coefficients calculated from Eqn.~(\ref{eq:StokesEinstein}) with a hydrodynamic diameter of twice the particle diameter, this yields times $\tau_\textrm d$ of 25~$\mu$s and 25~ms, respectively, for a displacement by the particle diameter.\\

Assuming a 1~MHz excitation rate and photon losses typical for state-of-the-art optics, a rough rule of single emitter localization and speed is given by \cite{PCCP_Jaime} $\sigma(\textrm{time, space}) = 1~\textrm{nm~Hz}^{-1/2}$. This means that at 40~kHz, the imaging rate corresponding to 25~$\mu$s, the localization precision is 200~nm, with only about 1 photon detected per image -- this is considerably larger than the length scale of interest. A different approach is needed to access timescales near or below 25~$\mu$s. Rather than suppressing the background by using fluorescence and detecting a wavelength other than the excitation wavelength, I decided to collect scattered light, which in suitable geometries is limited by photon input rather than the fluorescence process.\\

%

%

The totally internally reflected beam in the traditional \textsc{tirf} geometry is beneficial in that it does not lead to scattering in one of the liquid phases, and if the particles are added to the interface via one of the liquids, this means no background scattering from particles in the bulk liquid, i.e. a low background signal. With this sample illumination geometry, there are two typical imaging methods: i\textsc{scat} (signal composed of light scattered by sample and interferometric term between scattered and reflected light) and dark-field microscopy (scattered light only). In i\textsc{scat}, where the reflected illuminating beam is not removed from the collected signal, a larger signal is obtained compared to the dark-field geometry, and, if used with background subtraction techniques, it can achieve high localization precisions at high speeds. However, state-of-the-art i\textsc{scat} has not achieved wide-field illumination at imaging rates of several tens of kHz or higher.~\cite{iSCATrevSan} Besides, i\textsc{scat} is highly sensitive to fluctuations in sample backgrounds and the optical path. Meanwhile, experimental advances in dark-field microscopy \cite{Ueno,Berry}, in which the pure scattering term is detected, have also led to combinations of high localization precision and temporal resolution. This indicates the possibility of achieving the same spatiotemporal resolution with less sensitivity to background noise, which becomes important as the scattering intensity decreases with particle size and illumination. Because of the limited speed of i\textsc{scat} in wide-field applications and its higher sensitivity to background noise compared to dark-field geometries that suppress background noise \cite{Ueno}, I use dark-field microscopy in this work.\\

I characterize the state of the particle using the diffusion coefficient (as defined in Eq.~(\ref{eq:StokesEinstein})), the scattering intensity and the instantaneous diffusion coefficient, defined as $\tilde{D}(t) = (\vec{r}(t+\tau)-\vec{r}(t))^2/(4\tau)$. $t$ is a point in time, $\vec r$ is the position vector of the particle and $\tau$ is the time difference between measurements. The scattering intensity is converted from digital camera units to photoelectrons (which are related to photons scattered by the particle via the quantum efficiency of the camera, estimated to be 50\% from the specifications) using the photon transfer curve method (see Supporting Information for details). \cite{PTC_Li, Janesick} To the best of my knowledge, the relationship between scattering intensity and immersion angle in the oil phase is not known, and besides, the scattering intensity varies with lateral position as the illumination from the laser beam is not homogeneous. However, the scattering intensity is positively correlated with immersion depth within the range I observed. 
Furthermore, I expect the magnitude of fluctuations in intensity to be positively correlated with increased out-of-plane motion.

\section{Experimental}

\subsection{Imaging of particle dynamics}

\begin{figure}[H]
\begin{tabular}[t]{lll}
&
\multirow{2}{*}{\begin{subfigure}[t]{.65\textwidth}
\caption{}\label{fig:TIR-DF_bmpaths}
\includegraphics[width=\textwidth]{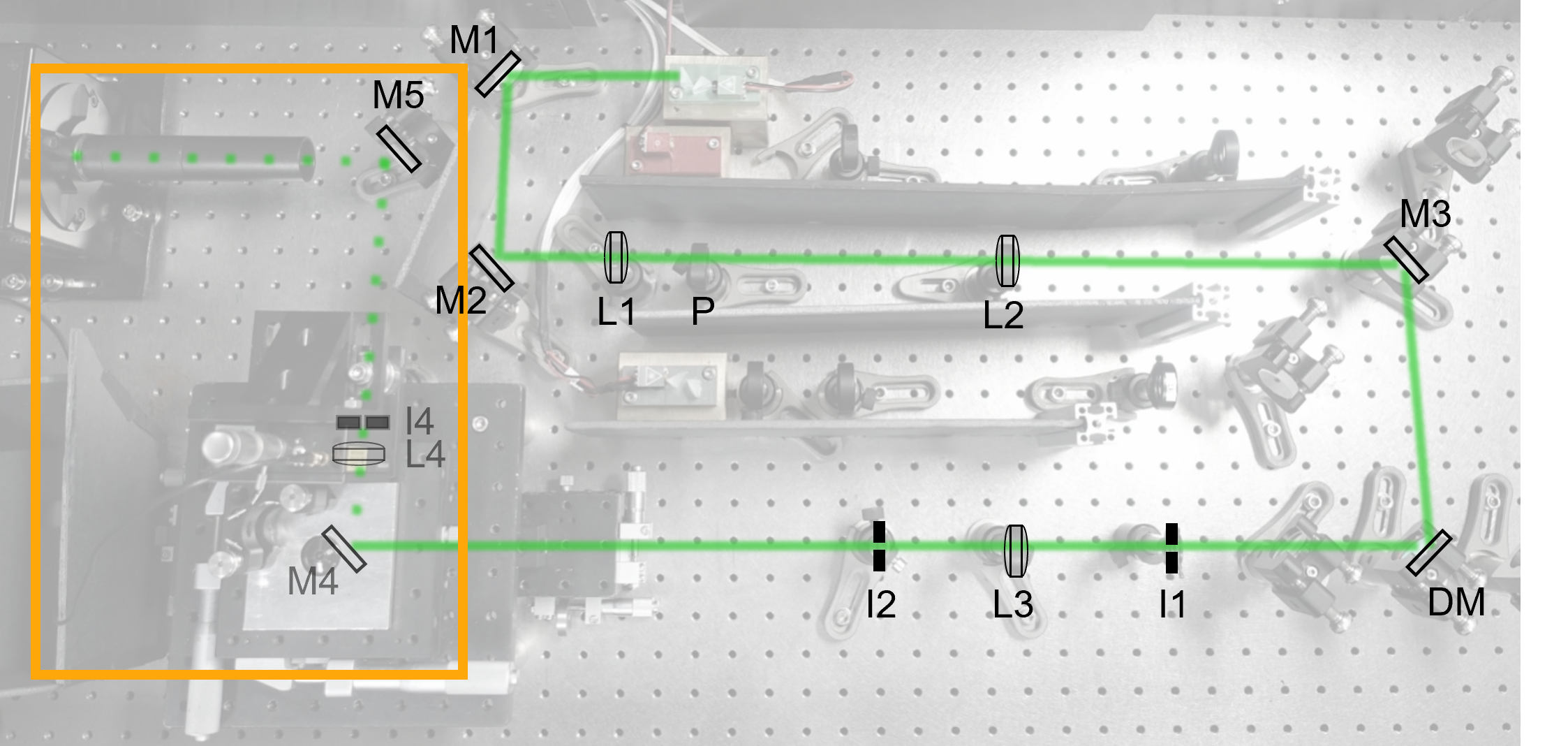} 
\end{subfigure}}
&
\begin{subfigure}[t]{.25\textwidth}
\caption{}\label{fig:TIR-DF_principle}
\includegraphics[height=2.1cm]{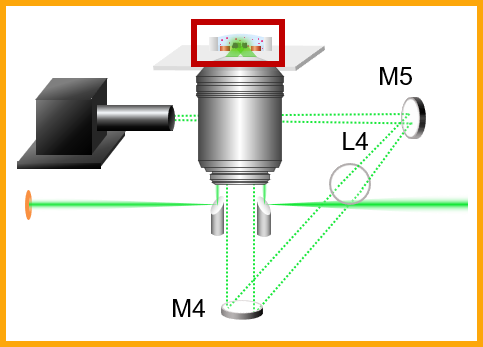}
\end{subfigure}
\\
&&
\begin{subfigure}[t]{.25\textwidth}
\vspace{0.2cm}
\caption{}
\includegraphics[height=2.1cm]{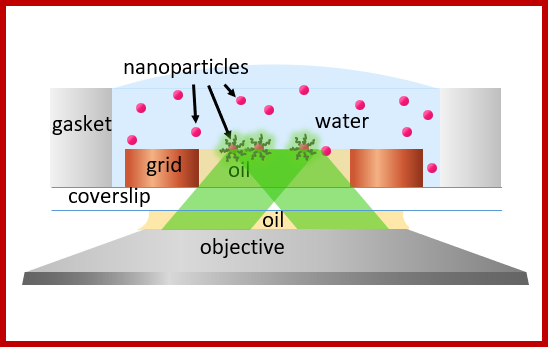}\\
\end{subfigure}
\end{tabular}\\
\vspace{0.1cm}

\begin{subfigure}[b]{\textwidth}
\vspace{-0.2cm}
\begin{tabular}{lcc}
&
\begin{subfigure}{.13\textwidth}\vspace{.3cm}\caption{}\label{fig:ADSvsskim}\includegraphics[height=1.1cm]{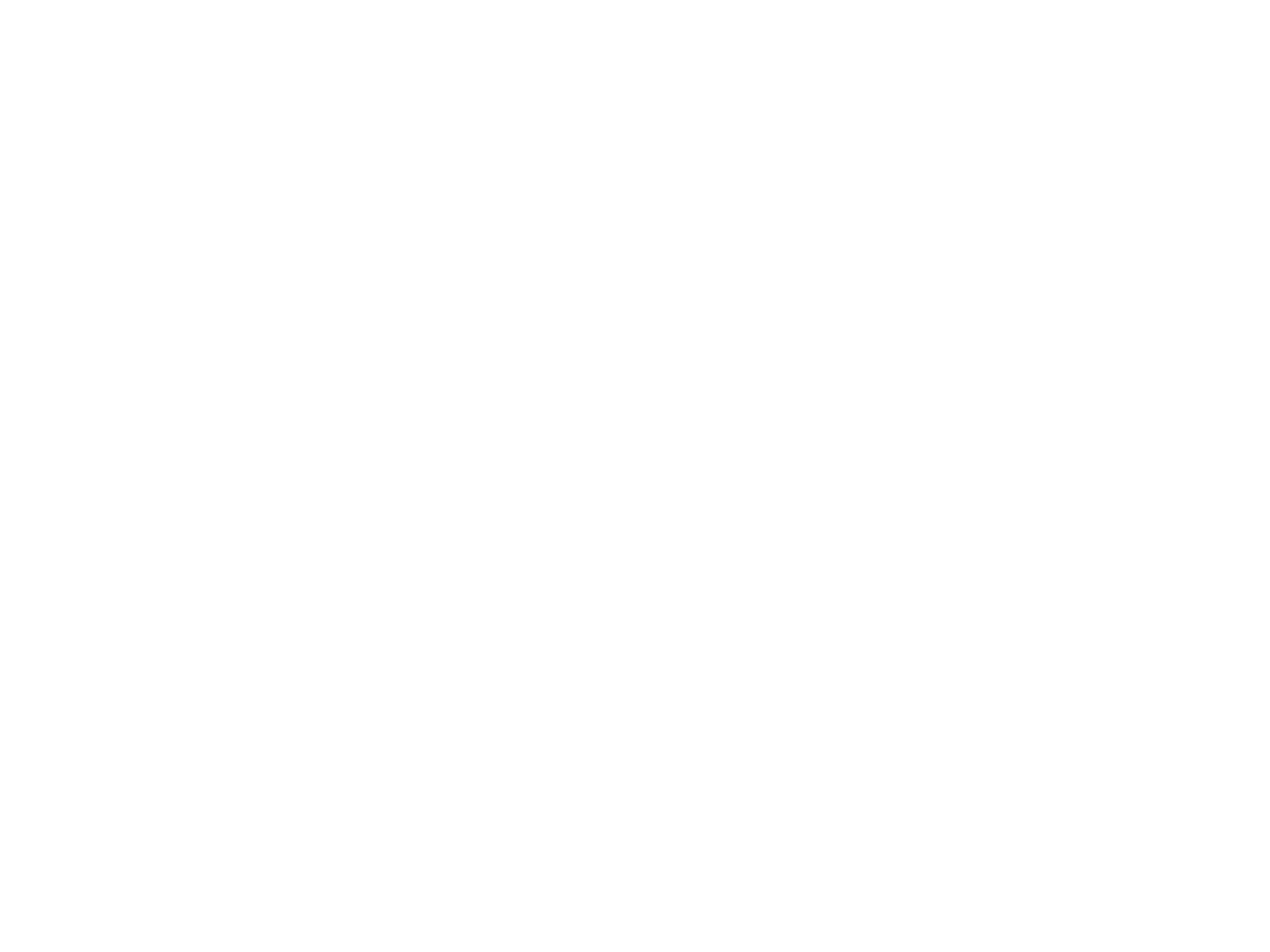}\end{subfigure}
\includegraphics[width=.8cm]{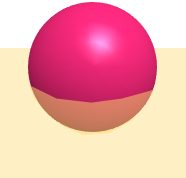}\begin{subfigure}{.13\textwidth}\includegraphics[height=1.3cm]{white}\end{subfigure}
&
\begin{subfigure}{.13\textwidth}\caption{}\label{fig:adsvsSKIM}\includegraphics[height=1cm]{white}\end{subfigure}
\includegraphics[width=.8cm]{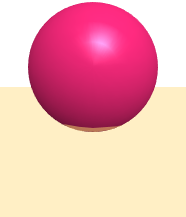}\begin{subfigure}{.13\textwidth}\includegraphics[height=1.3cm]{white}\end{subfigure}
\vspace{-.3cm}
\\

&
\includegraphics[width=4.5cm,trim={0 0 2.18cm 0},clip]{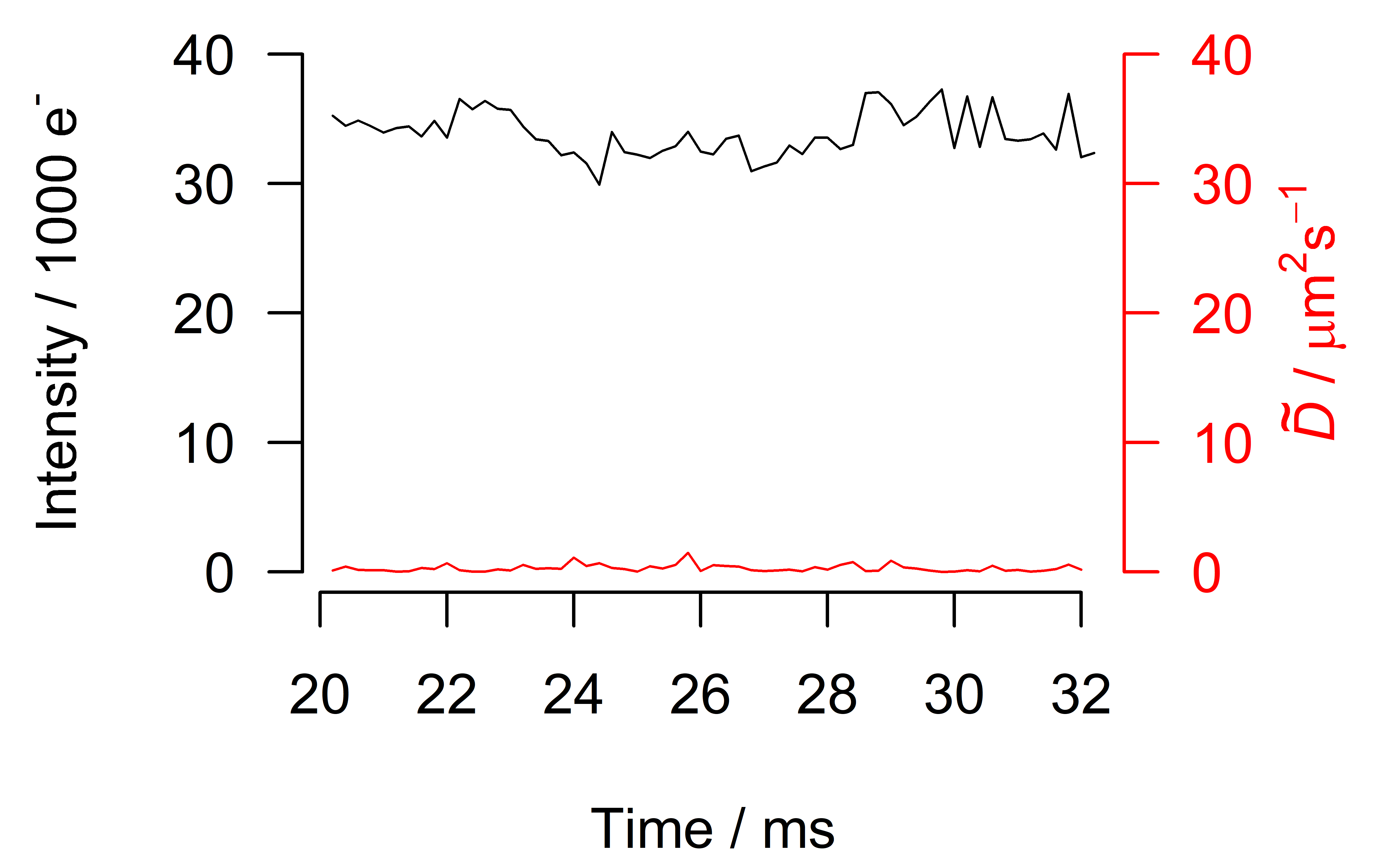}
\includegraphics[width=.05cm]{white}
&
\includegraphics[width=.2cm]{white}
\includegraphics[width=4.5cm,trim={2.335cm 0 0 0},clip]{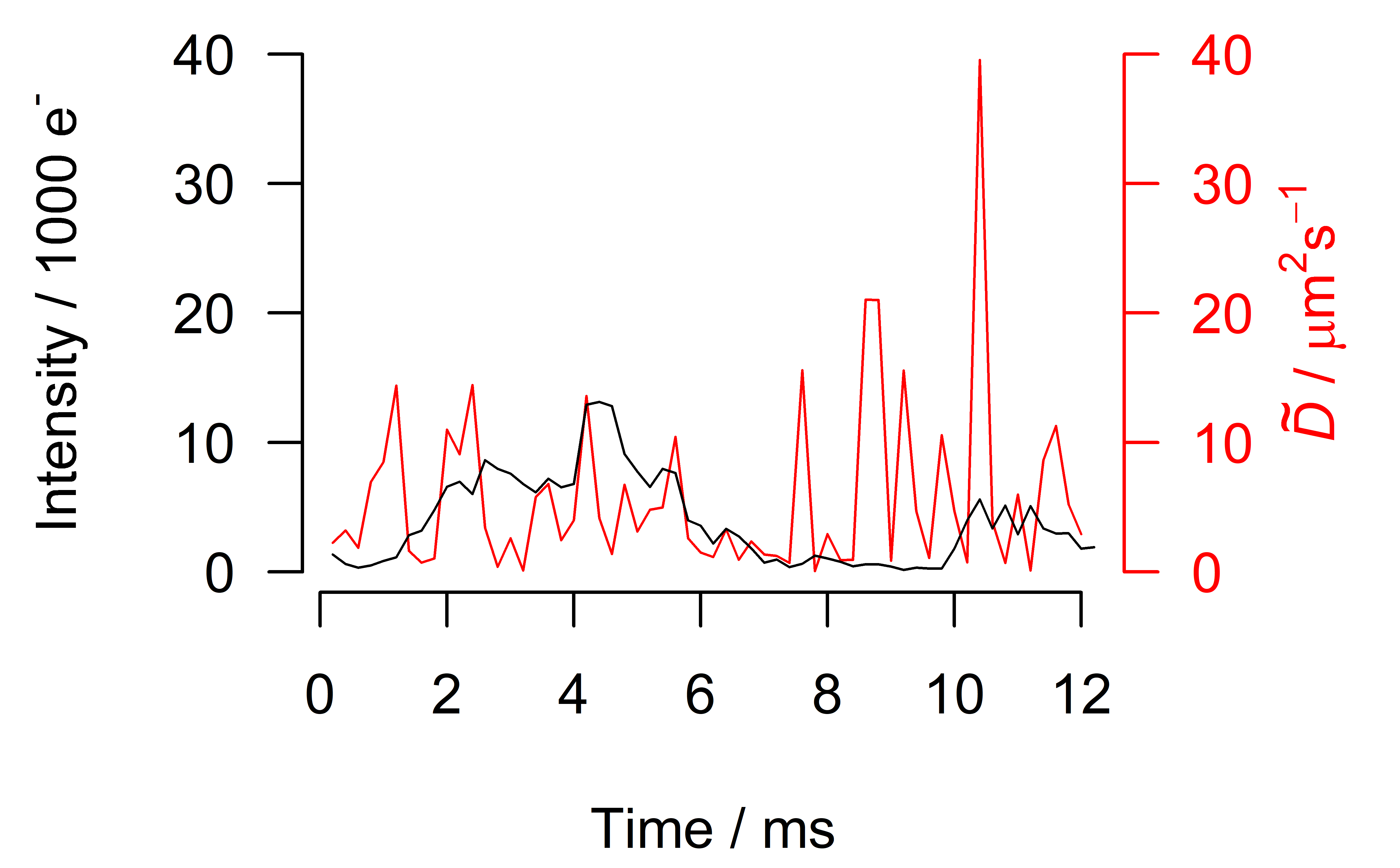}
\vspace{-.4cm}
\\

\includegraphics[scale=.1,valign=c,trim={8.27cm 0 0 0},clip]{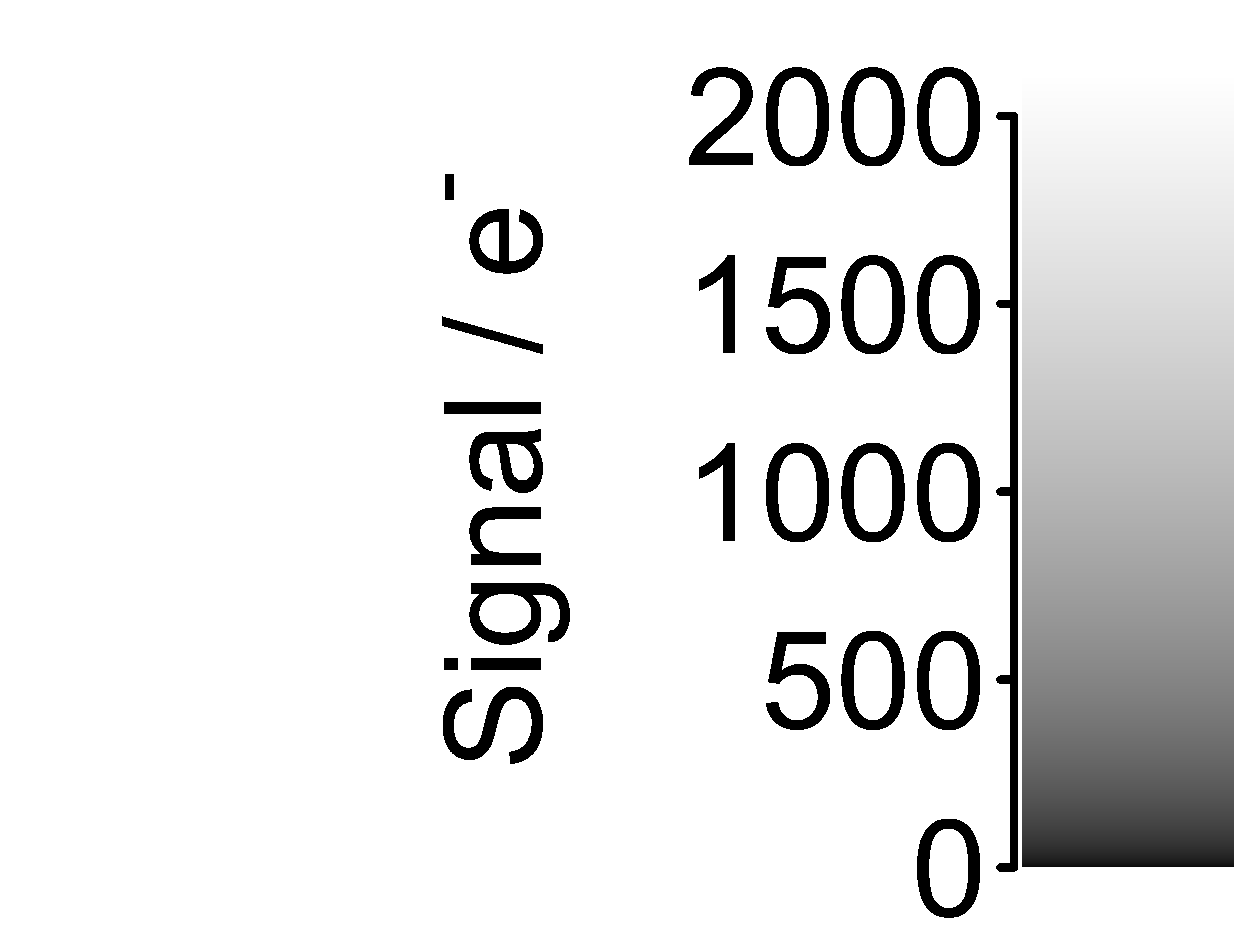}\hspace{-.3cm}
&
\includegraphics[width=.8cm,valign=c]{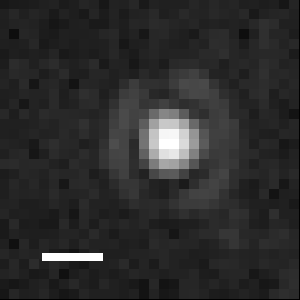}
\includegraphics[width=.8cm,valign=c]{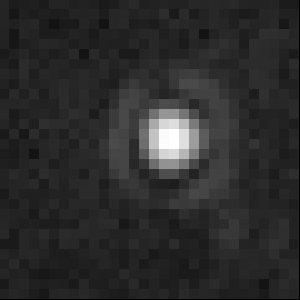}
\includegraphics[width=.8cm,valign=c]{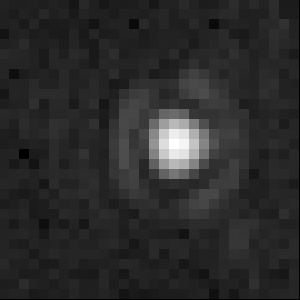}
\includegraphics[width=.8cm,valign=c]{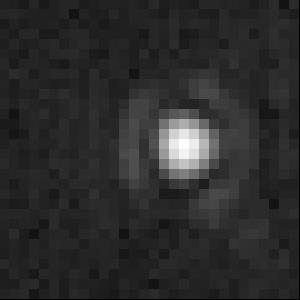}
\includegraphics[width=.8cm,valign=c]{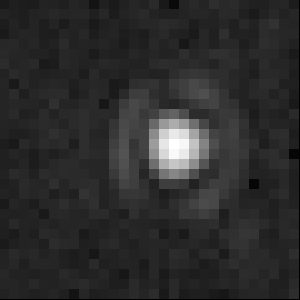}
&
\includegraphics[width=.8cm,valign=c]{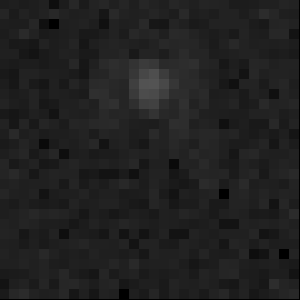}
\includegraphics[width=.8cm,valign=c]{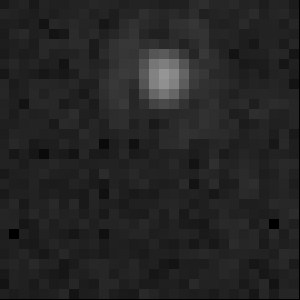}
\includegraphics[width=.8cm,valign=c]{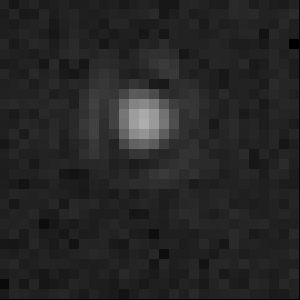}
\includegraphics[width=.8cm,valign=c]{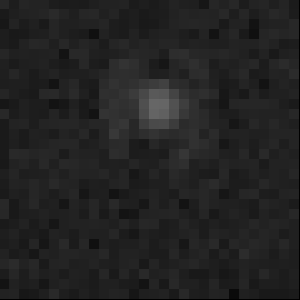}
\includegraphics[width=.8cm,valign=c]{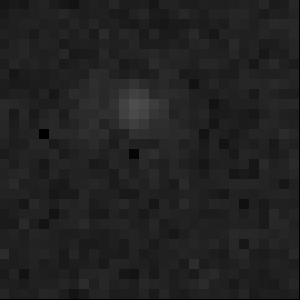}
\vspace{-.3cm}
\end{tabular}
\end{subfigure}
\caption{\textbf{Total-internal-reflection dark-field (TIR-DF) microscopy setup.} a)~Beam paths (in green) within the microscope; dotted where the light scattered by the sample is collected and focused onto the sensor of the camera. b)~Schematic close-up view of the sample area; marked in orange in a). c)~Schematic close-up view of an interface on a coverslip above the objective; marked in b) in red. The totally internally reflected beam is reflected by the oil-water interface and a small fraction of the incident light is scattered anisotropically by particles in the illuminated area. d)-e)~Intensity (black) and instantaneous diffusion coefficient (red) for a 40~nm gold particle at a MQ interface, at a sampling rate of 5~kHz. Intensity and instantaneous diffusion coefficient ($\tilde D$) scales are identical. d)~Adsorbed state and e)~a state in which the the particle is not fully adsorbed. Images of the same field of view with the same amplitude scaling are shown at intervals of 2~ms starting with the first frame of each trajectory segment. Scale bar: 500~nm.}
\label{fig:TIR-DF}
\end{figure}

The \textsc{tir-df} microscope, sample and some representative data I used to infer particle dynamics are shown in Fig.~\ref{fig:TIR-DF}.

The beam from a 1 W 520 nm diode laser (\textit{Lasertack LDM-520-1000-C}) 
is directed through a lens \textsc{l}1 ($f=50$~mm) followed by a 100~$\mu$m pinhole \textsc{p} in the focal plane of the lens. A 50~$\mu$m pinhole achieved an approximately Gaussian beam profile, 
however, this caused a power loss of almost 90\%. The 100~$\mu$m pinhole easily allowed 180~mW to pass, which is likely to be at the limit of what the objective can tolerate based on past experience in the group. The beam profile is not as close to Gaussian as with the smaller pinhole, however, and this is visible in that when a particle attached to glass is moved across the field of view, there are several intensity maxima with the change in lateral position. For many of the earlier experiments, a single-mode optical fibre was used instead, replacing the beam path between \textsc{l}1 and \textsc{l}2 in Fig.~\ref{fig:TIR-DF_bmpaths}. The beam is re-collimated using another lens \textsc{l}2 ($f=200$~mm), and the beam is reflected by a 
mirror and a long-pass dichroic mirror DM that allows the beam from the 635~nm laser (\textit{Lasertack LDM-635-200}) to pass but reflects the 520~nm beam. It then passes through another long-pass dichroic mirror included to allow coupling in of a 445~nm laser (\textit{Lasertack LDM-445-2000-CC}), and is focused by a 400~mm achromatic lens \textsc{l}3 and optionally adjusted with two irises \textsc{i}1 and \textsc{i}2 before being reflected into the \textit{Olympus Plapon 60XO} objective with \textsc{na}~1.42 
and working distance 150~$\mu$m by a micromirror. The beam is collimated by the objective and directed at the interface at an angle achieved by lateral displacement of the micromirror from the centre of the objective, leading to total internal reflection (see Fig.~\ref{fig:TIR-DF}). A second rod mirror is used to remove the reflected beam (shown in Fig.~\ref{fig:TIR-DF}b, but not a), and the position of this beam can be used to infer the distance between the interface and the objective. The light scattered by the sample is collected by the objective, transmitted through an iris \textsc{i}3 (not shown) to remove stray reflections, focused by a 400~mm achromatic lens \textsc{l}4 and directed through an iris \textsc{i}4 onto the camera chip of the \textit{Photron FASTCAM Mini UX100 type 800KM - 16GB} camera (pixel size 10~$\mu$m). One pixel in the resulting image corresponds to 75~nm in this setup, chosen in relation to the expected spatial extent of the signal in the image to optimize localization precision \cite{LPtheo}. A typical field of view was 640 pixels wide, i.e. 48~$\mu$m; the other dimension varied with the imaging speed and desired movie length.\\

Particles that are more immersed in the oil phase have a stronger scattering signal and diffuse more slowly within the interfacial plane. Particles at shallower immersion depths are more exposed to the water phase, which has a lower viscosity and, according to Eqn.~(\ref{eq:StokesEinstein}), allows for faster diffusion.

\subsection{Interface sample preparation}

The interface must be stable for a sufficiently long time period to allow observation, and possibly equilibration, as there may be flows immediately after the interface is established. To visualize the interface with a \textsc{tir-df} microscope, the interface must be brought within the working distance of the objective, which is typically less than 300~$\mu$m for oil-immersion objectives. Background scattering must be minimized. Contaminants are undesirable both in terms of making the experiment more difficult due to their scattering, and in that contaminants could have a wide range of effects \cite{Kralchevsky_book} on the experiment that would alter the results in unpredictable and adverse ways. The interfacial tension between the liquids must be such that the adsorbed state is energetically favored, which must be considered when choosing the liquids. Planarity is desired to avoid effects characteristic to the curvature of the interface \cite{Stebe_curvature}, which would be hard to characterize experimentally because the exact curvature would be difficult to reproduce. Small isotropic curvatures, however, are likely to be negligible. A study \cite{Du2012_droplet3d} found that curvatures of droplets where the radius of curvature was at least an order of magnitude larger than the radius of curvature of the particle had no effect on the diffusion coefficient of particles at the interface between the droplets and the surrounding fluid. Both phases having a low viscosity makes it possible to observe particle motion in the underdamped regime. Finally, it is desirable for particles to not be too strongly bound to allow the observation of dynamics perpendicular to the interface.\\

I developed an interface sample protocol adapted from ref.~\cite{JACS_Meli}. A copper aperture is placed on a coverslip. A fine pipette tip attached to the tip of a Pasteur pipette is used to transfer a small amount of centrifuged oil to the centre of the copper grid. A silicone gasket with a circular hole is placed over the copper ring, and finally, the water phase containing the nanoparticles is carefully added on top of the oil with a pipette (approximately 20~$\mu$L). I aim for a slightly overfilled grid, which results in a stable and flat or slightly overfilled oil-water interface. A cover is placed on the sample during data acquisition to minimize the background signal from ambient light as well as evaporation and the influence of air currents. Particles may start adsorbing to the interface within 20 minutes after adding the aqueous phase, especially for particles in \DDM{} solution, but depending on the interface sample it may take more than an hour for the first particles to be observed. As a rule, I observed smaller particles to land sooner after the preparation of the interface, and sooner in \DDM{}-oil interfaces than in water-oil interfaces (see Fig.~\ref{fig:timeinterface}). This could be due to different diffusion properties in the water phase: Small particles are expected to diffuse faster (see Eqn.~(\ref{eq:StokesEinstein})), and the adsorption of \DDM{} molecules to the particles could alter their hydrodynamic radius.

\section{Results}

\begin{figure}[H]
\begin{subfigure}[t]{.42\textwidth}
\caption{}\label{fig:ads_MQ_overlay}
\includegraphics[width=6.4cm]{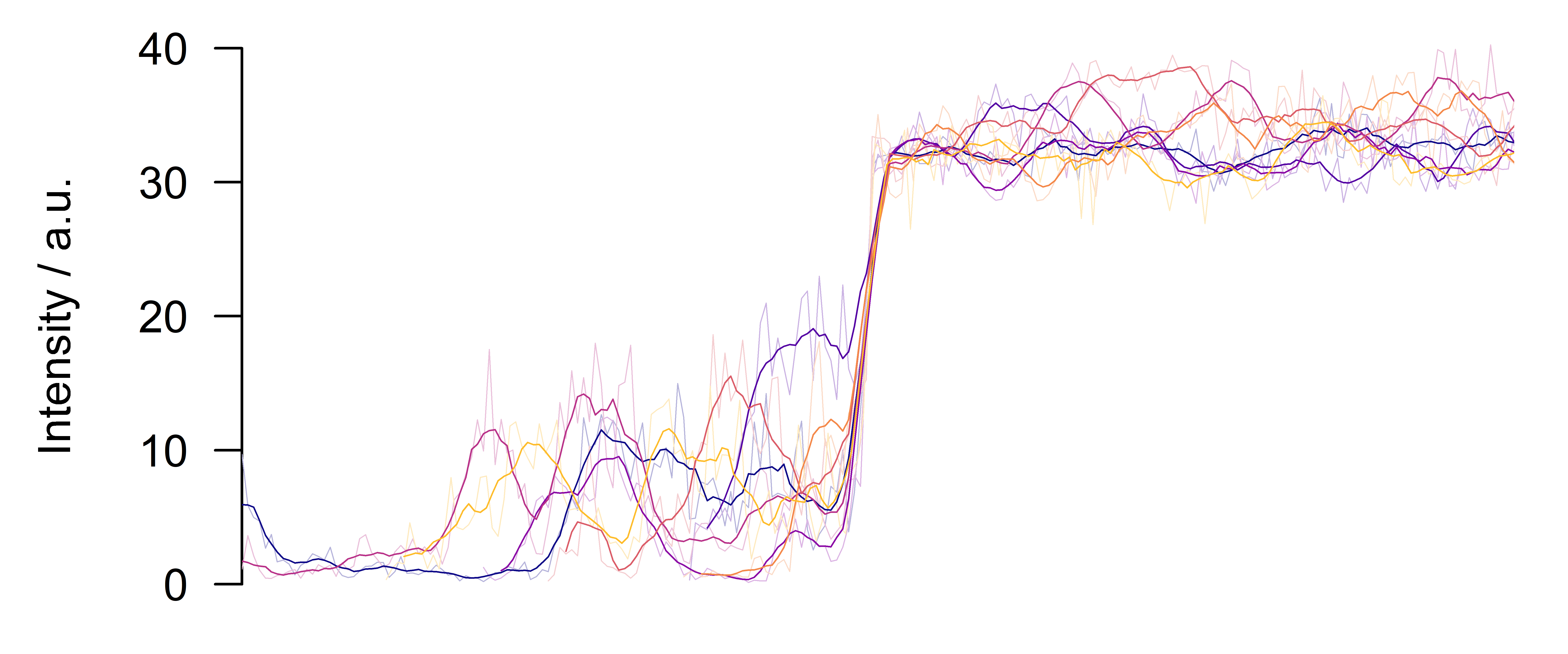}\\
\includegraphics[width=6.4cm]{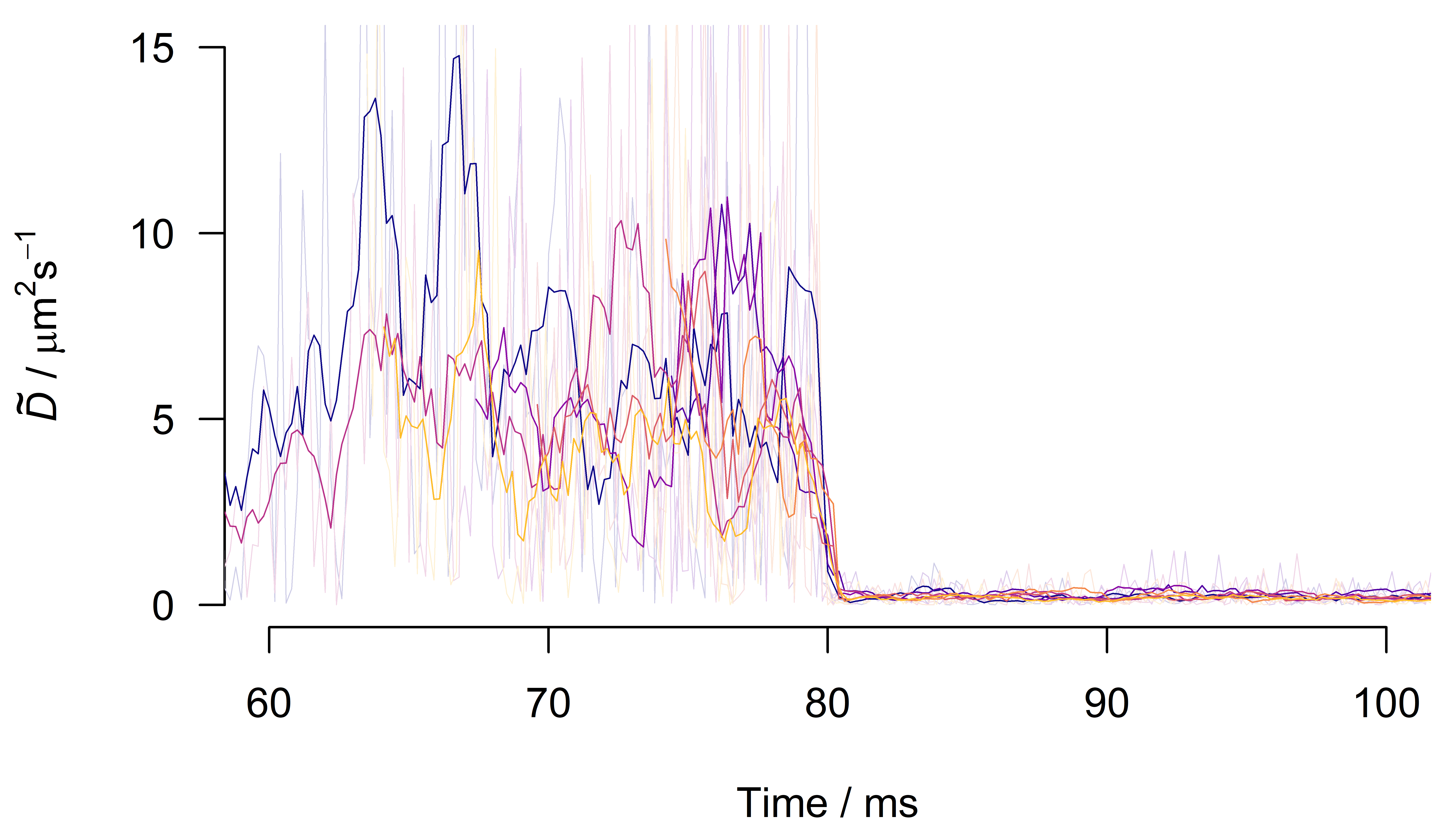}
\end{subfigure}
\begin{subfigure}[t]{.56\textwidth}
\caption{}\label{fig:traces}
\includegraphics[width=8.9cm,trim={0 .15cm 0 0},clip]{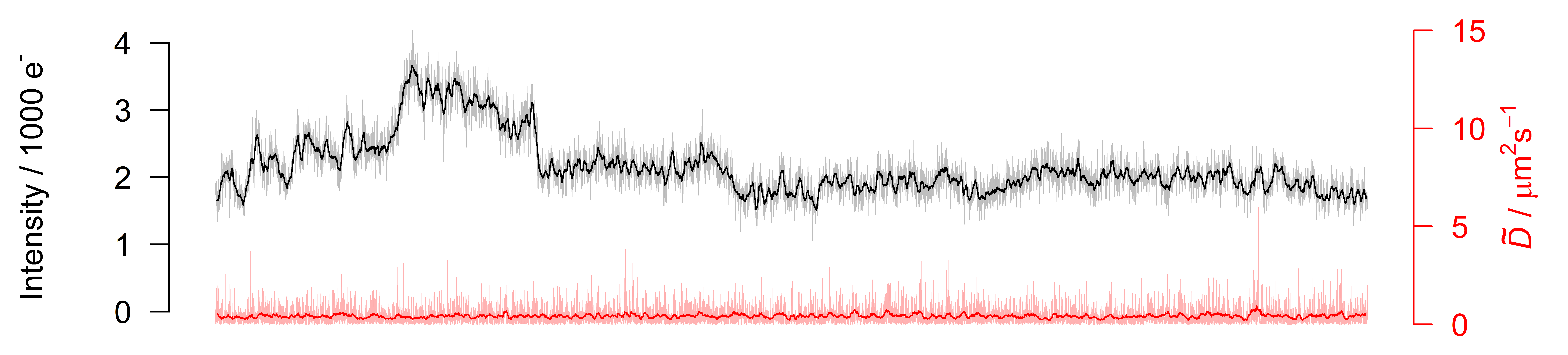}
\includegraphics[width=8.9cm,trim={0 0.9cm 0 4.45cm},clip]{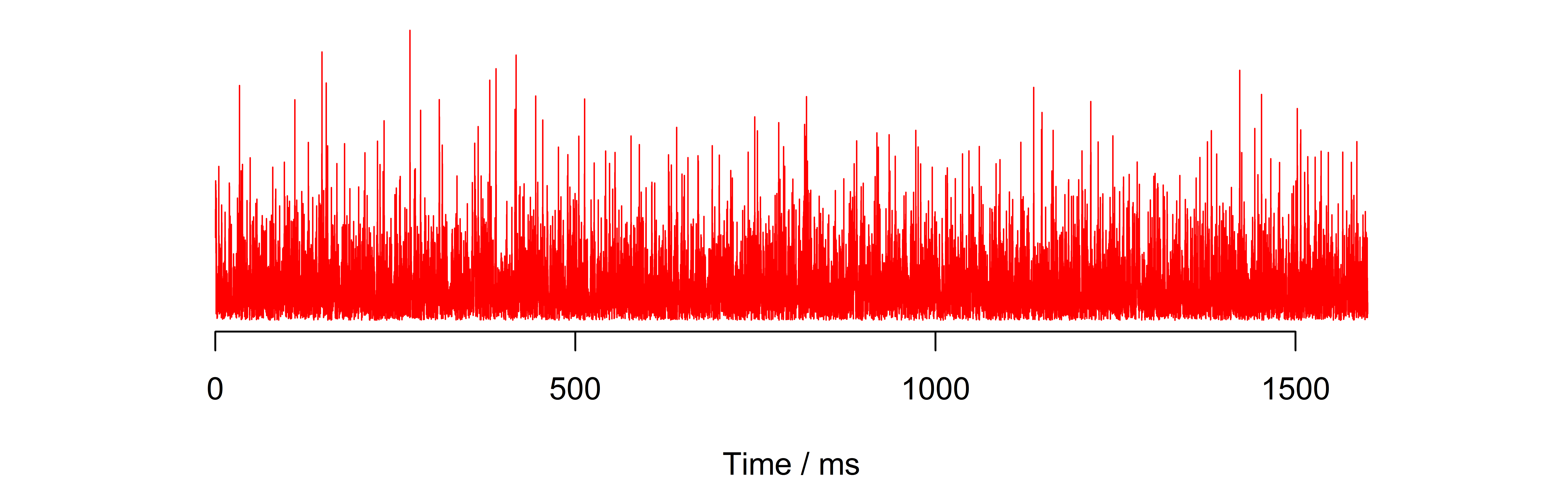}\\ 
\includegraphics[width=8.9cm,trim={0 0.9cm 0 0},clip]{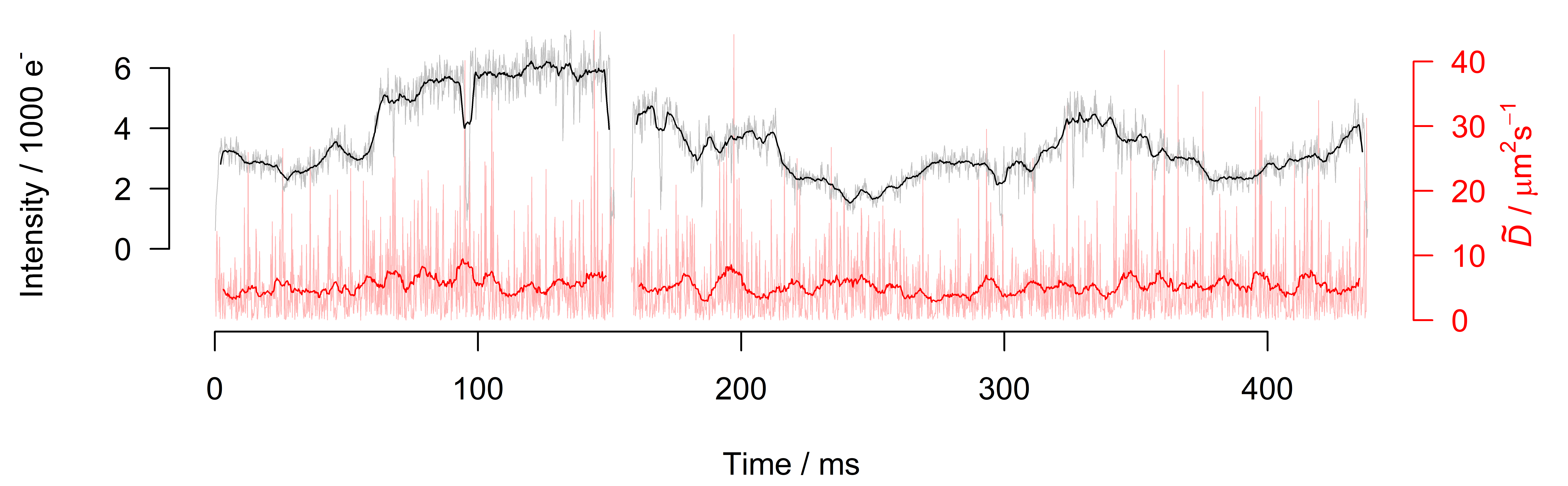} 
\includegraphics[width=8.9cm,trim={0 0.18cm 0 0},clip]{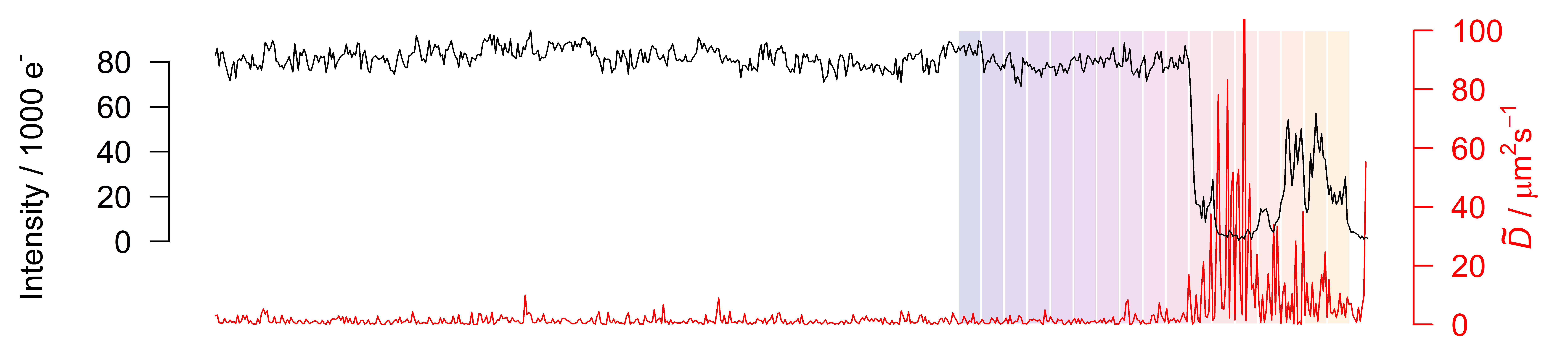}\\ 
\includegraphics[width=8.9cm,trim={0 0 0 4.49cm},clip]{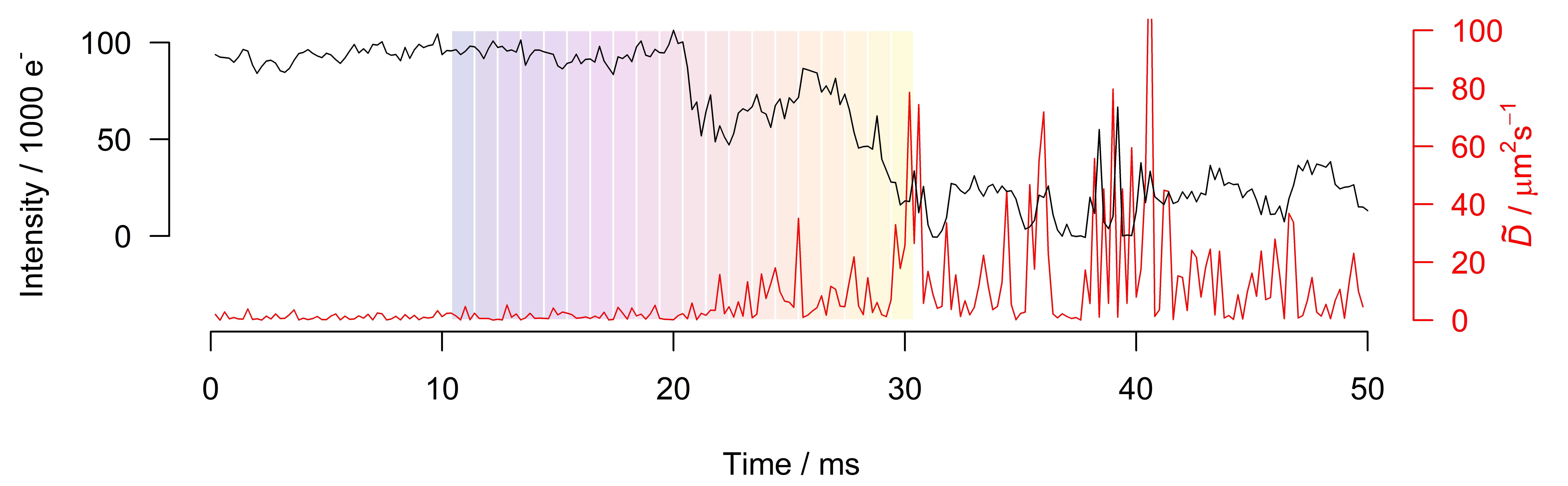}\\ 
\begin{picture}(0cm,0cm)
\put(1.5cm,2.23cm){\includegraphics[width=2.36cm,trim={0 2.3cm 0 0},clip]{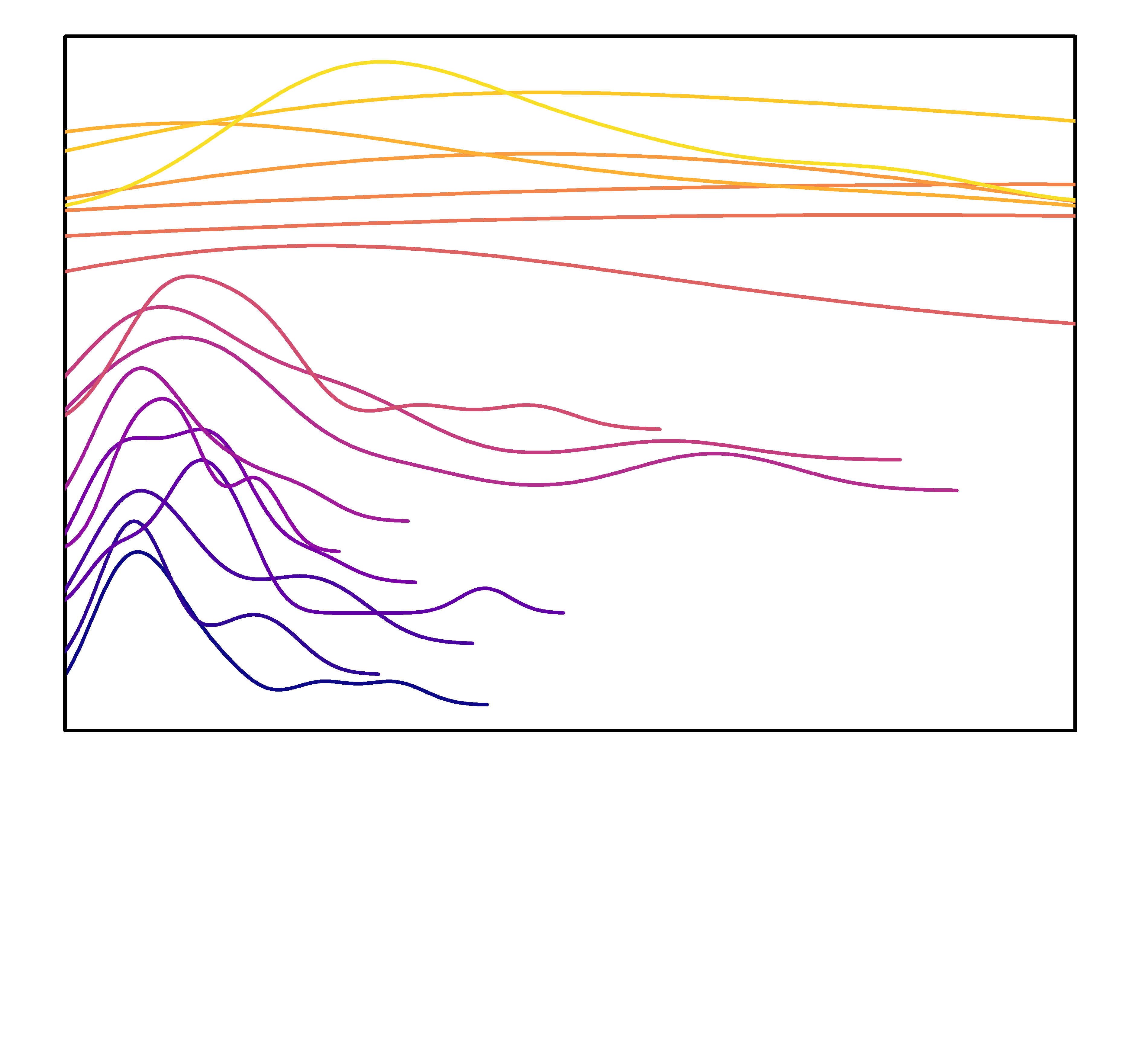}}
\put(1.5cm,1.55cm){\includegraphics[width=2.36cm,trim={0 0 0 5.2cm},clip]{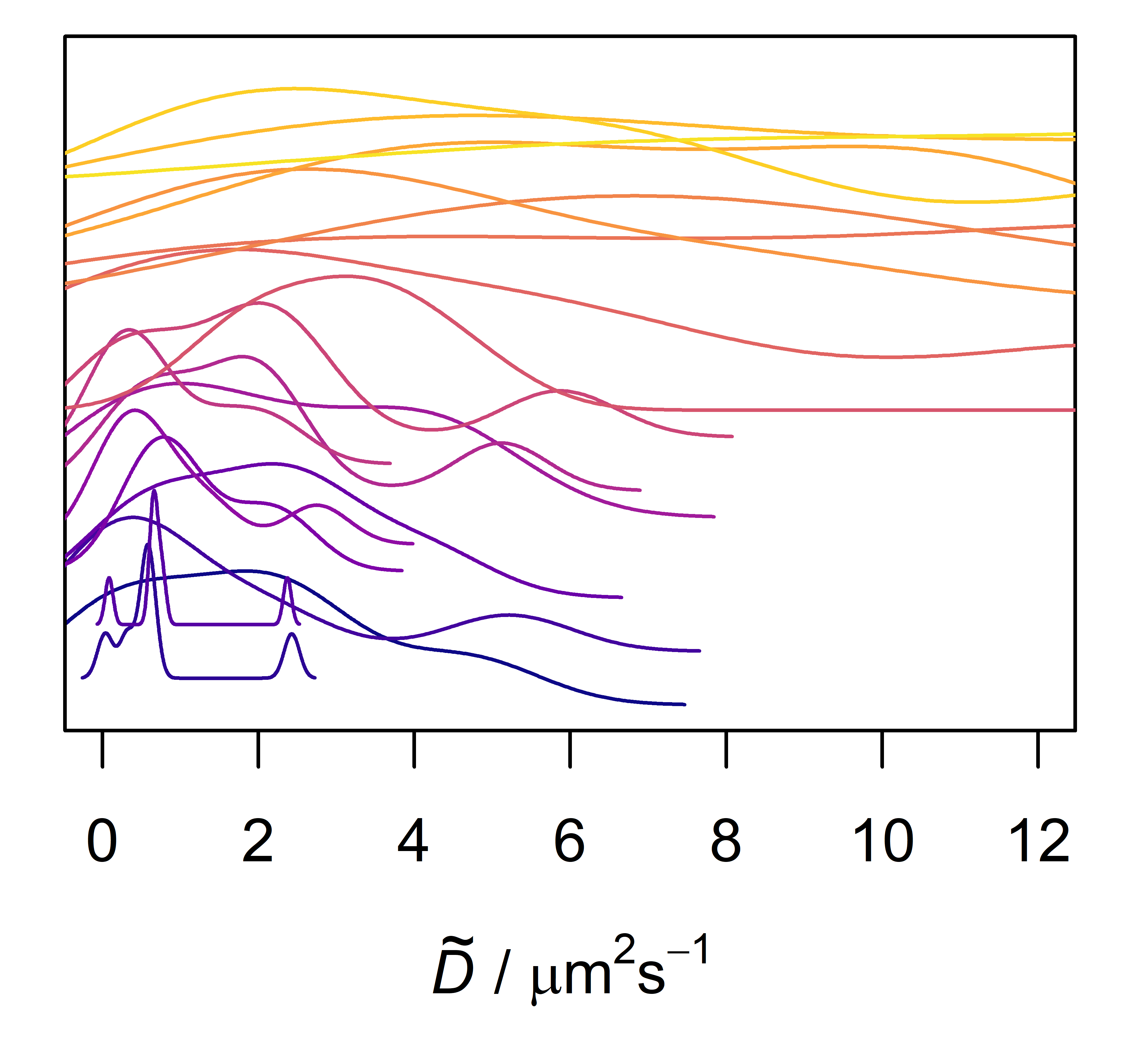}} 
\end{picture}
\vspace{-0.5cm}
\end{subfigure}
\caption{\textbf{Representative examples of particle dynamics observed.} Where there are darker and lighter lines in the same color, darker lines represent smoothed values. a) Intensity and instantaneous diffusion coefficient traces of seven different 40~nm gold particles, shown in different colours, during adsorption events at surfactant-free interfaces, sampled at 5~kHz. The trajectories are shifted in time and the intensities are scaled so that the first frame with an intensity within the adsorbed range and the intensity at this time are the same for all trajectories. b) Top to bottom: i) 20~nm particle at a \DDM{} interface, ii) skimming the oil phase (briefly adsorbing without lateral motion slowing down to the typical adsorbed value, then desorbing again), iii) 40~nm particle desorbing (inset shows instantaneous diffusion coefficient).}
\label{fig:behaviors}
\end{figure}

Details of the data set used are given in Tab.~\ref{tbl:data_agg_breakdown}, and some representative behaviors are shown in Fig.~\ref{fig:behaviors}. At surfactant-free (\MQ{}) interfaces, adsorption is rapid ($\lesssim100$~$\mu$s) and no relaxation to equilibrium was observed (see Fig.~\ref{fig:ads_MQ_overlay}). No desorption events were observed. At \DDM{} interfaces, both adsorption and desorption events were recorded, indicating a significantly lower interfacial stabilization energy. This is expected from the addition of a surfactant. However, larger particles were found to adsorb to \DDM{} interfaces significantly less frequently than to \MQ{} interfaces, with 80~nm particles hardly adsorbing at all --- despite larger particles usually stabilizing interfaces more than smaller ones, as they remove a larger interfacial area given the same contact angle. Some particles at \DDM{} interfaces exhibit dynamics that I did not observe at \MQ{} interfaces (see Fig.~\ref{fig:traces}). Partially adsorbed states are common, and the faster-diffusing of these states often feature brief flights through the water phase, with duration increasing with particle size. Some trajectories contain abrupt transitions between immersion states.\\

\begin{table}[H]
\caption{Overview of the dataset used for ensemble analyses. The number indicates the particle diameter in nm; \MQ{} and \DDM{} refer to the interface type (without and with added surfactant, respectively). Trajectories shorter than 1000 frames are not included.}
\begin{tabular}{l c c r l}
Group			& Interfaces		& {Trajectories}	& {Data points} 	& {Sampling rates / kHz}	\\
\toprule
20 \MQ{}			& 5			& 416			& \ph\ph4\ph678\ph579	 	& {2.5, 5, 12.5}		\\
\hspace{0.4cm} \DDM{}	& 4			& 641			& \ph\ph5\ph452\ph645	 	& {2.5, 5, 12.5}		\\
40 \MQ{}			& 5			& 652			& 10\ph646\ph135 	& {2.5, 5, 12.5, 25, 50}	\\
\hspace{0.4cm} \DDM{}	& 3			& 208			& \ph1\ph903\ph460		& {2.5, 5, 12.5, 25}	\\
80 \MQ{}			& 5			& 271			& \ph4\ph778\ph569	 	& {2.5, 5, 12.5, 25, 50}	\\
\hspace{0.4cm} \DDM{}	& 1			& 43				& \ph\ph\ph\ph\ph897\ph555	 	& {\hspace{.73cm}5, 12.5}\\
\end{tabular}
\label{tbl:data_agg_breakdown}
\end{table}

\begin{figure}[H]
\begin{tabular}{ll}
\begin{subfigure}{.44\textwidth}
\caption{}\label{fig:data_D_MQ}
\includegraphics[width=5 cm]{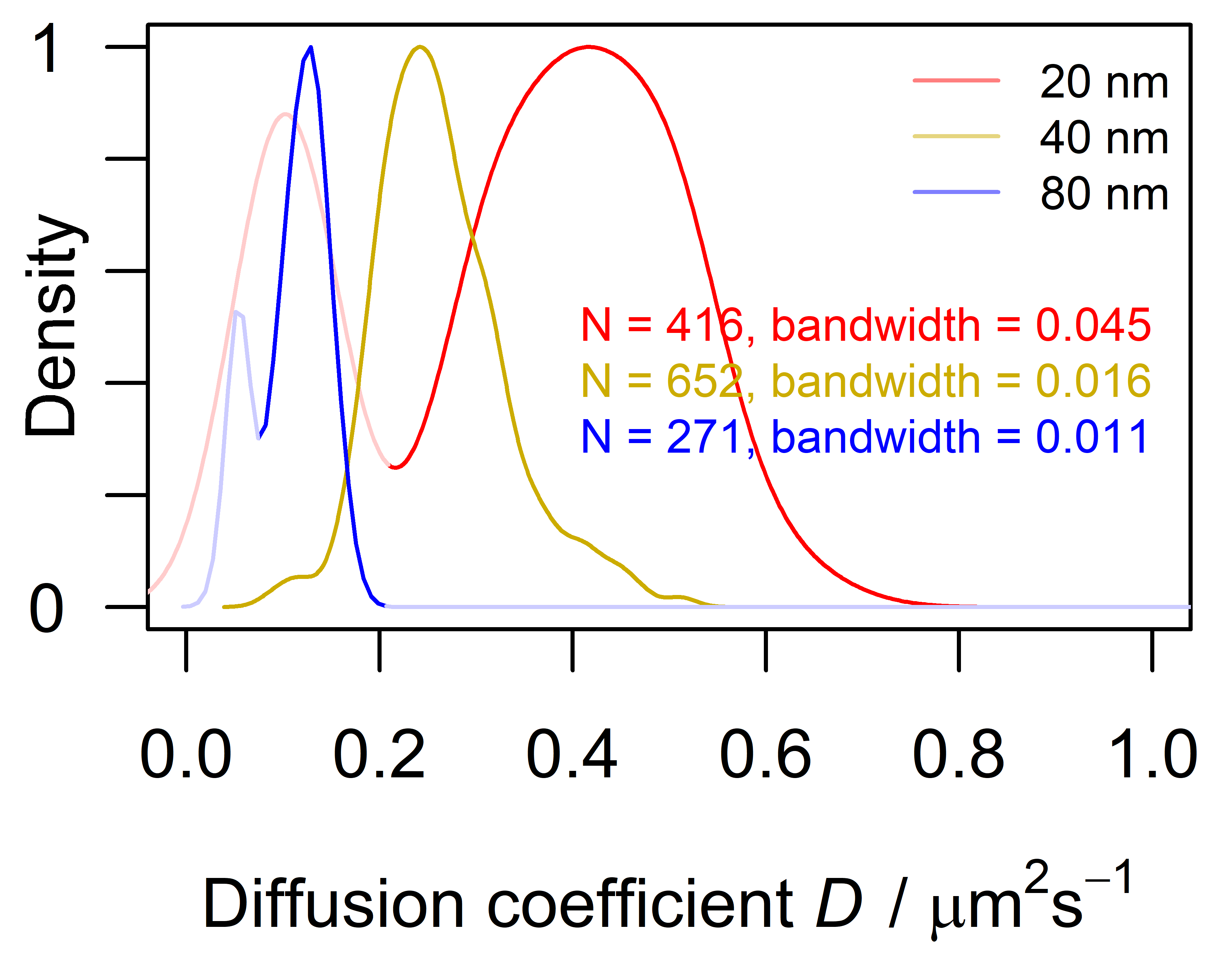}
\end{subfigure}&
\begin{subfigure}{.44\textwidth}
\caption{}\label{fig:data_D_DDM}
\includegraphics[width=5 cm]{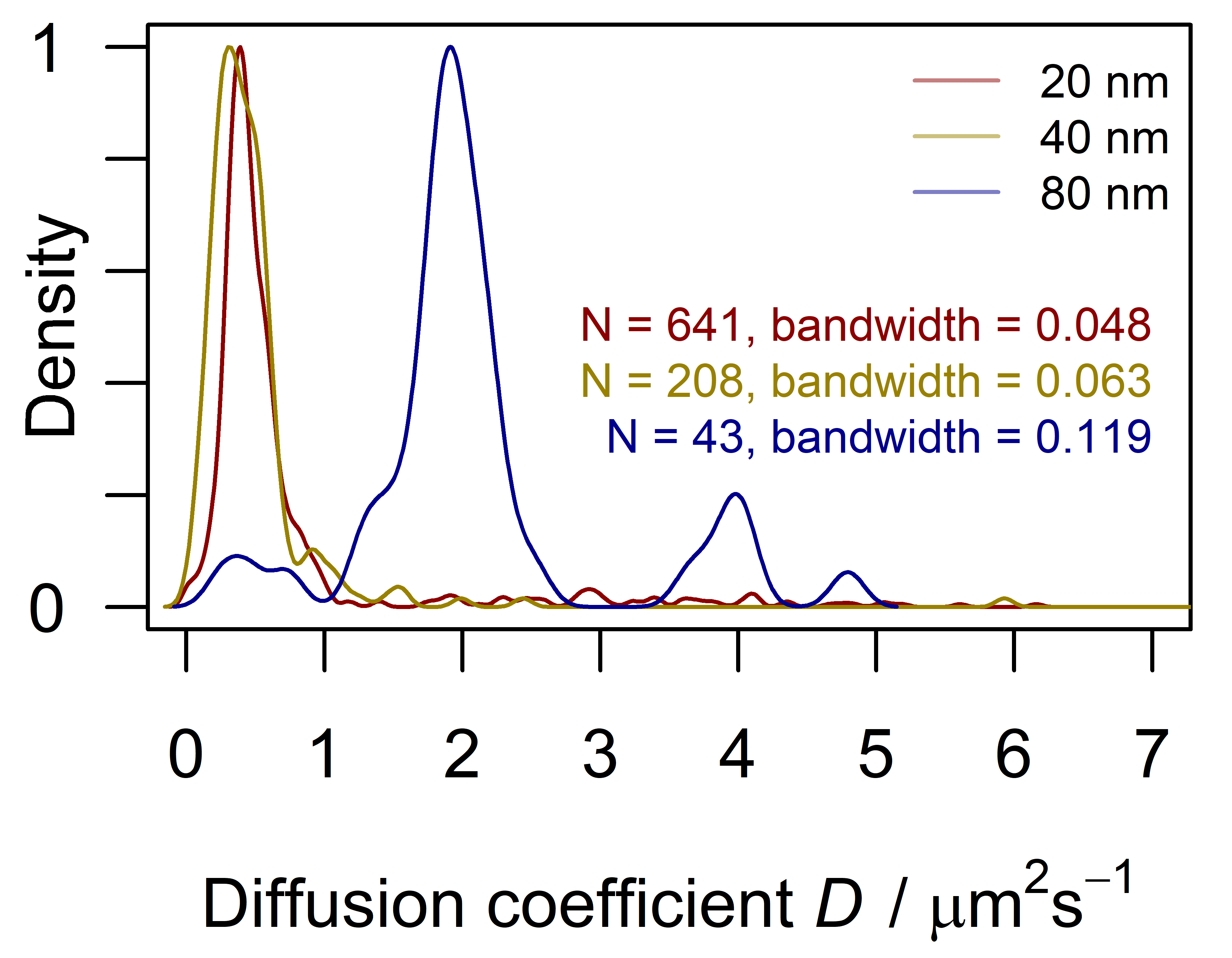}
\end{subfigure}\\
\vspace{.1cm}
\begin{subfigure}[b]{.44\textwidth}
\caption{}\label{fig:data_D_MQ_szscld}
\includegraphics[width=5 cm]{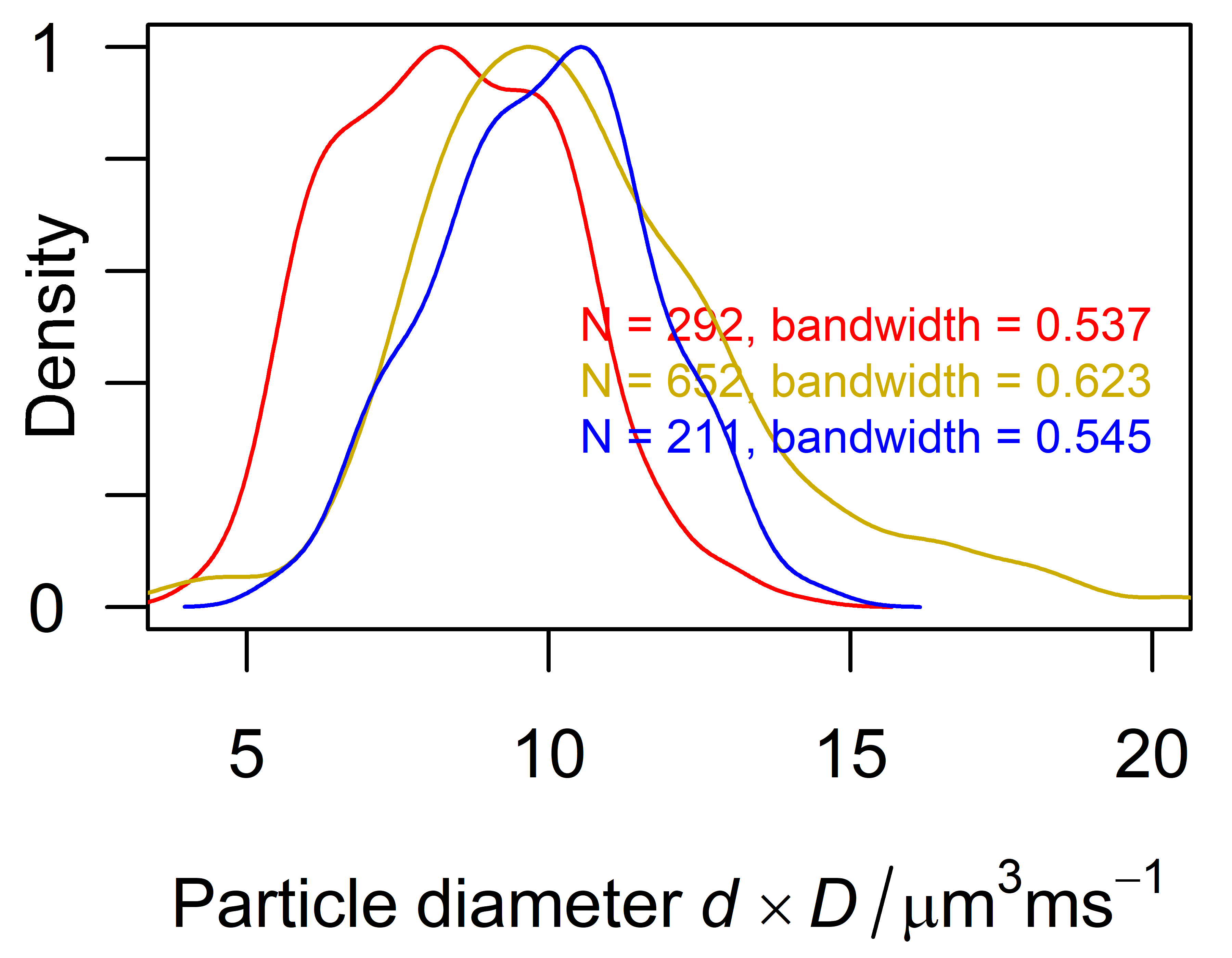}
\end{subfigure}&
\begin{subfigure}[b]{.25\textwidth}
\caption{}\label{fig:data_D_rb}
\includegraphics[width=3.3cm]{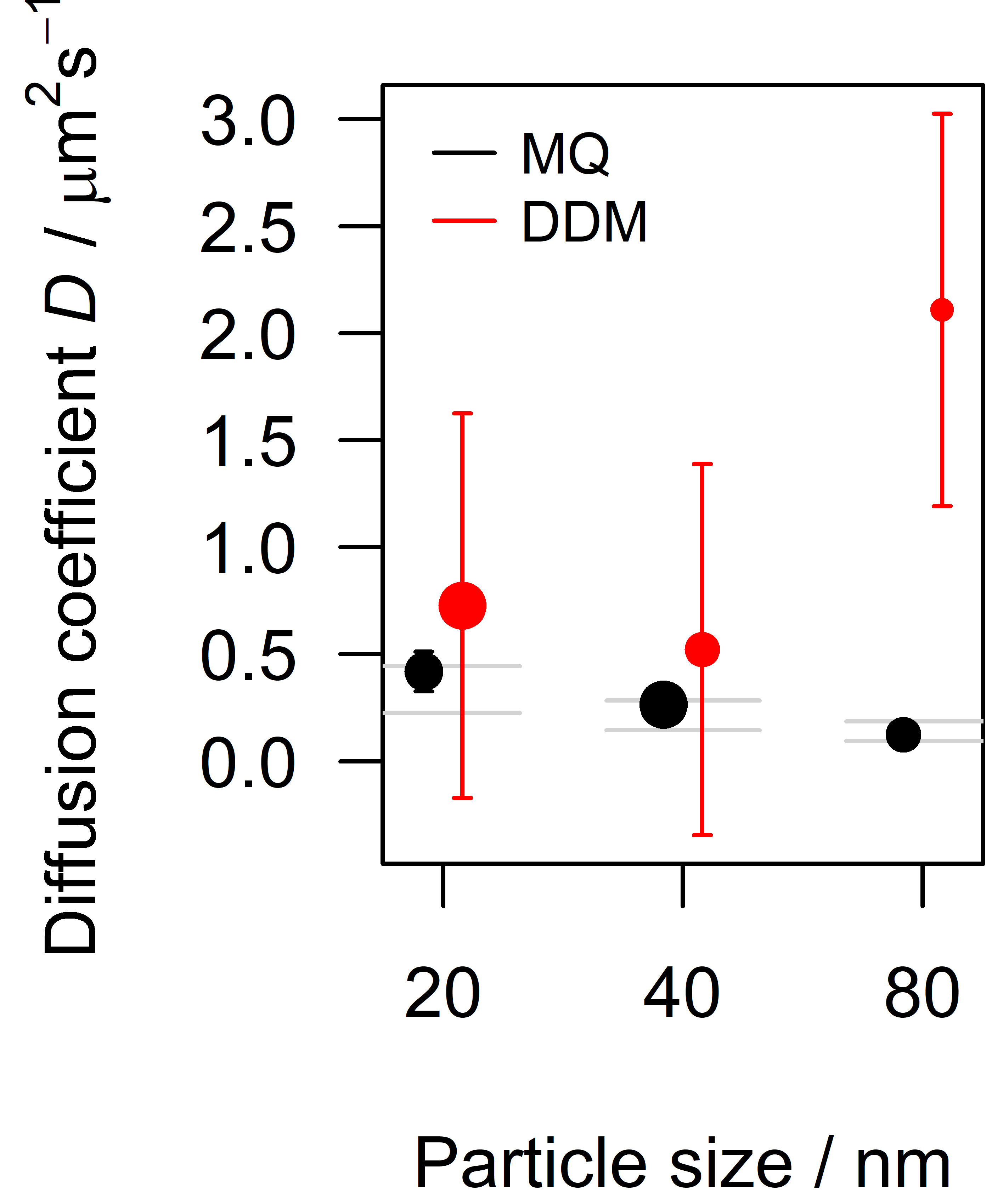}
\end{subfigure}
\end{tabular}
\caption{\textbf{Diffusion coefficients of gold particles at oil-water interfaces.} Density distributions at surfactant-free (\MQ{}, a) and surfactant-laden (\DDM{}, b) interfaces. c) Distribution of size-scaled diffusion coefficients at \MQ{} interfaces, with shoulders and outliers removed in the 20 and 80~nm distributions from a) (fainter lines in a)). The shoulders are thought to be due to contamination or aggregation, as there is a sharp division between them and the main peaks (see Fig.~\ref{fig:difcoefshoulders}). d) Distribution maxima and standard deviations within each particle size and interface type. The size of the points in the graph scales with the number of trajectories $n$ as $\propto n^{0.3}$. Grey lines represent diffusion coefficient values corresponding to theoretical contact angles of 0 (fully immersed in the oil) and 90$\degree$, respectively. The numbers given here assume the Stokes radius provided by the manufacturer (see section~\ref{sec:materials}).
}
\label{fig:data_D}
\end{figure}

The diffusion coefficient distributions of gold particles at interfaces with and without surfactant are shown in Fig.~\ref{fig:data_D}. At surfactant-free interfaces, I observed the diffusion coefficient distribution expected from the Stokes-Einstein equation (Eqn.~(\ref{eq:StokesEinstein})): The maximum for 20$\;$ nm particles is at approximately two times that of 40$\;$nm particles, which is double that of 80$\;$nm particles. The 20$\;$nm distribution has peaks at lower diffusion coefficients -- this might be a result of out-of-plane motion accounting for a part of the thermal energy of the particle, which is not significant in the larger particles. At interfaces with added \DDM, the diffusivity trend looks completely different: The 20 and 40$\;$nm particles' diffusion coefficients are similar to the values at surfactant-free interfaces, but with a much higher variance, and the 80$\;$nm particles are much more diffusive than any other particle and interface combination investigated.\\


\begin{figure}[H]
\begin{tabular}{ll}
\begin{subfigure}{.48\textwidth}
\caption{}
\includegraphics[width=5.1cm]{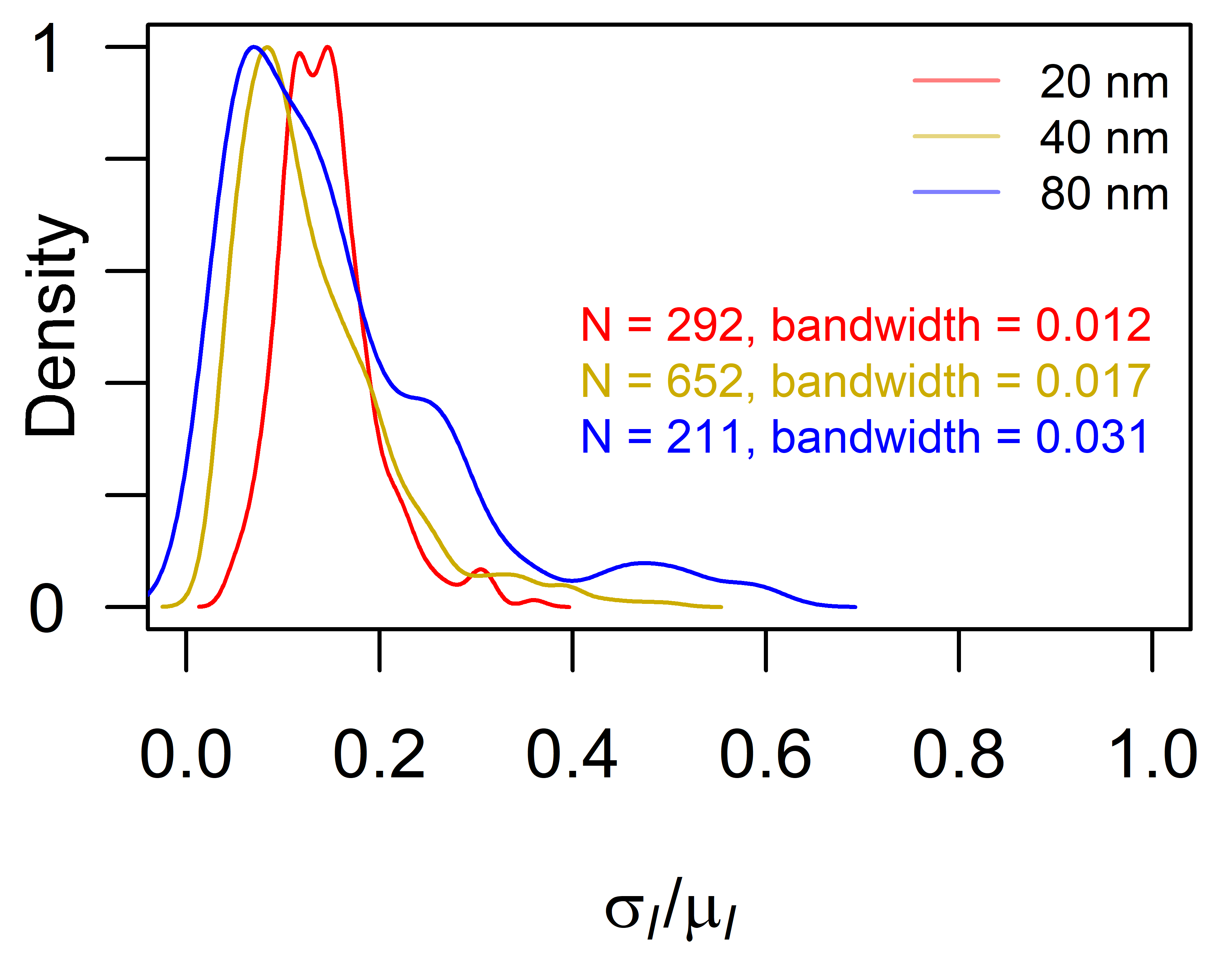}
\end{subfigure}&
\begin{subfigure}{.48\textwidth}
\caption{}\label{fig:int_DDM}
\includegraphics[width=5.1cm]{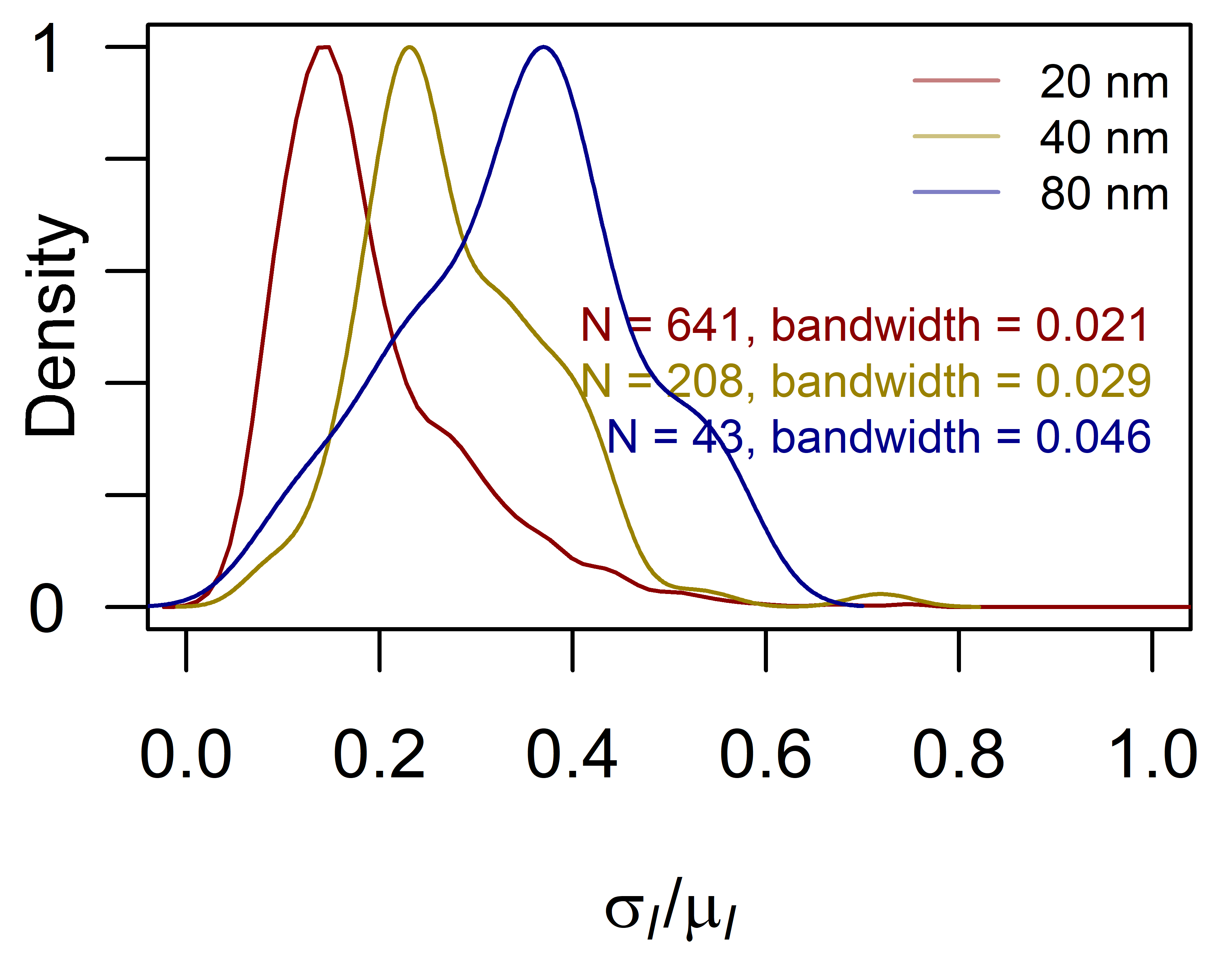}
\end{subfigure}\\
\begin{subfigure}{.05\textwidth}
\includegraphics[width=1cm]{white}
\end{subfigure}
\begin{subfigure}{.35\textwidth}
\caption{}
\includegraphics[width=3.5cm]{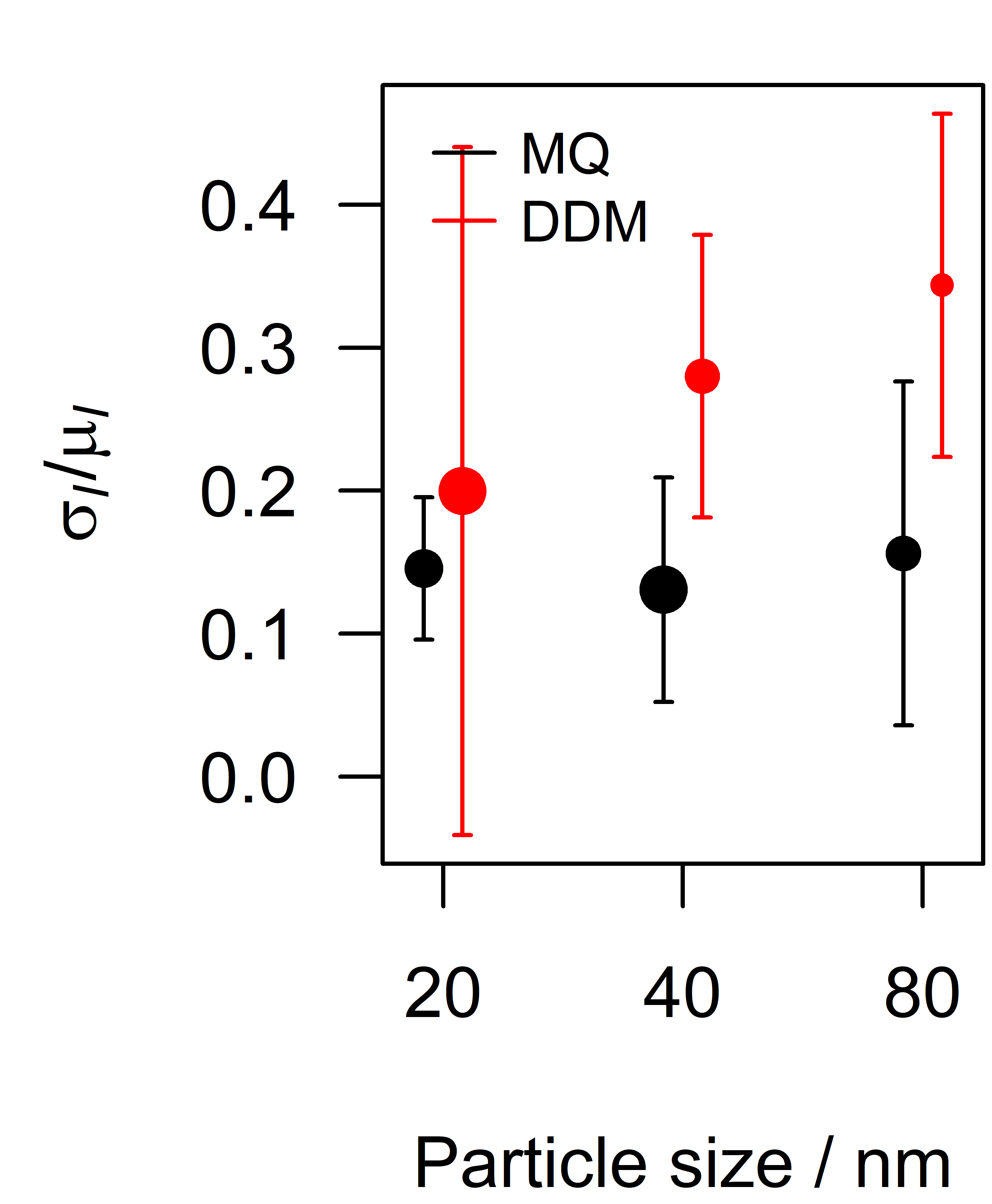}
\end{subfigure}&
\begin{subfigure}{.48\textwidth}
\caption{}
\includegraphics[width=5.1cm]{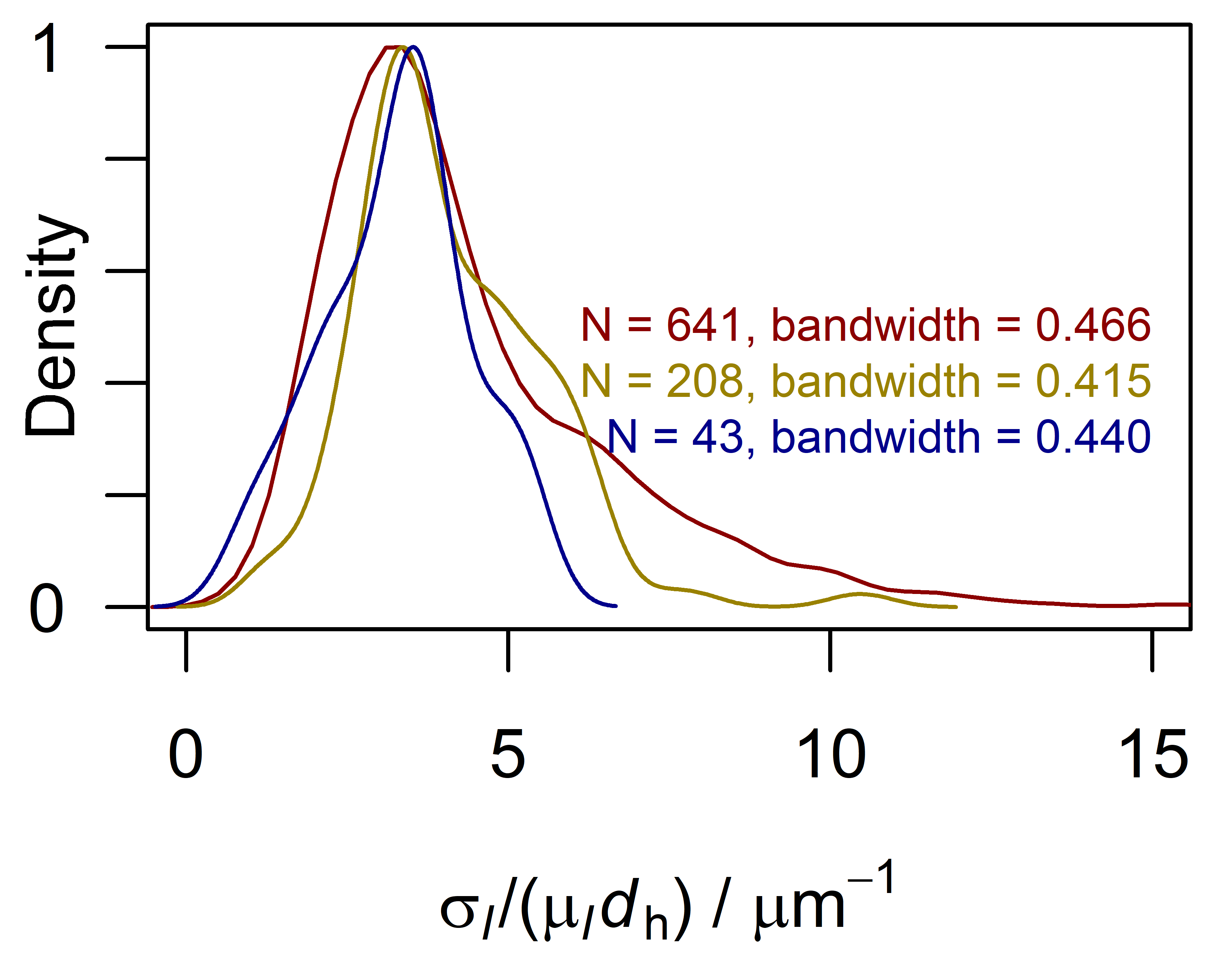}
\end{subfigure}
\end{tabular}
\caption{\textbf{Qualitative out-of-plane motion based on scattering intensity fluctuations.} Standard deviation $\sigma_I$ over a trajectory relative to trajectory mean $\mu_I$, a)~at \MQ{} interfaces and b)~at \DDM{} interfaces. c)~Distribution maxima of $\sigma_I/\mu_I$ as a function of particle diameter. d)~$\sigma_I/\mu_I$ at \DDM{} interfaces scaled with the hydrodynamic diameter $d_\textrm h$ of the particle.}
\label{fig:int}
\end{figure}

To understand this initially surprising result, I investigated the out-of-plane motion using the scattering intensity (Fig.~\ref{fig:int}). I find that at surfactant-free interfaces, the mean-scaled variation in the intensity is distributed similarly between different particle sizes. At \DDM{} interfaces, however, there are three distinctly different maxima in the distribution for the three different particle sizes, with fluctuations increasing with particle diameter.

This indicates increased out-of-plane motion with increased particle size; another reversal of the expected size trend. The combination of rarity of adsorption events and increased in-plane diffusivity and out-of-plane motion in large particles leads me to propose a confinement effect by surfactant molecules: Dense coverage of the interface with surfactant molecules both reduces the thermodynamic driving force for particle adsorption by lowering the interfacial tension, and may pose a kinetic and statistical barrier to adsorption, with the added requirement for surfactant molecules to move aside during the adsorption process, or for particles to encounter a sufficiently large vacant area. Larger particles are most strongly affected by this because they are less likely to fit into existing gaps in the surfactant layer and, because of their lower curvature, exert less lateral force on surfactant particles they could otherwise displace. The shallow immersion resulting from an incomplete adsorption would lead to the observed faster diffusion, as a larger proportion of the particle is immersed in the less viscous water phase. In addition, locally varying resistance from surfactant molecules may lead to confined in-plane diffusion. A diffusion analysis (Fig.~\ref{fig:data_alpha}) supports the confinement hypothesis.\\

Interestingly, the lateral diffusion in 20 and 40~nm particles at \DDM{} is not much faster than at \MQ{} interfaces despite the expected lower contact angle and the observed reduced scattering intensity. A reason for this could be increased motion perpendicular to the interface, made energetically accessible by lowering the interfacial tension. The resulting increase in the dimensionality of the diffusion distributes the same thermal energy over more degrees of freedom, decreasing the in-plane motion in favor of out-of-plane motion.\\

\section{Conclusions}

This study shows that there are new and unexpected interfacial particle dynamics to be explored in systems with an added surfactant, underlining how advantageously chosen and carefully optimized imaging methods can shed light on otherwise inaccessible dynamics. My findings demonstrate that there are unexplored, but technologically relevant dynamics in systems including mixtures of particles and surfactants, with surfactants considerably altering the particle dynamics. The reversed size trend in adsorption stability could open an avenue to replacing larger particles at interfaces with smaller ones, in contrast to the typical behavior in which larger particles displace smaller ones in systems with mixed particle sizes.\\

Varying the surfactant concentration, adding salt and otherwise expanding the range of experimental conditions would shed more light on the mechanisms behind these dynamics, and is likely to provide insights that will enable more control in applications. Investigating the dynamics of mixtures of different particle sizes would similarly be likely to yield new insights, with no changes to the experimental setup required. In addition, it would be possible to probe the potential energy landscape of the interface using methods very similar to ours, which could be used to establish whether the anomalously slow diffusion observed by me and others at surfactant-free interfaces can be explained by the effects of the line tension or other causes, and quantitatively investigate the dynamics observed at \DDM{} interfaces. In future experiments, one could capture the $z$ position of the particle by adding a cylindrical lens to the imaging system \cite{Huang_astigmatism}, causing the optical signal of particles to change as a function of the displacement from the interface. This three-dimensional spatial information could contribute greatly to the characterization of the relationship between the structure of the composite system and its properties, deepening scientific understanding and opening new pathways in the design of functional materials.


\section{Acknowledgements}
This work was carried out within a DPhil program in the laboratory of Prof. Philipp Kukura at the University of Oxford, United Kingdom. Prof. Kukura and Dr. Adar Sonn-Segev conceptualized the original research questions that this work aimed to answer. Dr. Xuanhui Meng provided helpful advice on how to optimize the microscopy setup.

\section{Supporting Information}

\subsection{Camera characterization: Quantifiying the signal}
\label{sec:camera}


To quantify the measured signal, the relationship between the digital numbers read from the camera and the number of photons arriving or the number of photoelectrons generated at the detector must be determined. The data on the specifications sheet provided by camera manufacturers may be an optimal or average value, so for quantitative measurements, it is often better to characterize the camera using the method of the photon transfer curve \cite{PTC_Li}. A more in-depth theoretical background can be found in Ref.~\cite{Janesick}.

The stochastic nature of the process of photons arriving at the detector makes it possible to convert the digital number displayed by the camera to the number of photoelectrons detected. The number of photoelectrons generated over the integration time of the camera chip (exposure time) follows a Poisson distribution, meaning that the expectation values for the signal and the variance of the signal are the same:

\begin{equation}
{\sigma_\textrm{shot}}^2 = S,
\label{eq:shot_noise}
\end{equation}

where $S$ is the signal arising purely from electrons generated by arriving photons and $\sigma_\textrm{shot}$ is its standard deviation. $\sigma_\textrm{shot}$ is called shot noise. Because the above equation holds in units of photoelectrons, the digital number to electron conversion factor, also known as the gain, can be obtained by dividing the signal in digital units by the squared shot noise in digital units:

\begin{equation}
C' =  \frac{S_\textrm{DN}}{{\sigma_\textrm{shot,DN}}^2},
\label{eq:DNtoelectron}
\end{equation}

resulting in the digital number to photon conversion factor

\begin{equation}
C = \Phi_\lambda \frac{S_\textrm{DN}}{{\sigma_\textrm{shot,DN}}^2},
\label{eq:DNtophoton}
\end{equation}

where $\Phi_\lambda$ is the quantum efficiency (photoelectron to photon ratio) at photon wavelength $\lambda$.

With an ideal, noise-free camera, shot noise would be the only contribution to the noise, and the conversion factor could be determined by measuring the standard deviation of the digital signal as a function of the digital signal. In practice, there are sources of noise related to the experimental setup that must be taken into account. For signals that do not saturate the detector, the total noise of a low-light camera can be approximated by \cite{PTC_Li}

\begin{equation}
{\sigma_\textrm{total}}^2 = {\sigma_\textrm{read}}^2 + {\sigma_\textrm{shot}}^2 + {\sigma_\textrm{FP}}^2,
\label{eq:camera_noise}
\end{equation}

where the individual contributions are from read noise, shot noise and fixed-pattern noise. The read noise is the minimum operating noise of the camera and does not change with the signal. Fixed-pattern noise is due to differences between pixels. In CMOS cameras, the latter is a large contribution to noise measured over an image; however, it can be eliminated by treating individual pixels rather than the entire camera chip, which is what is done here to avoid overestimating the shot noise. When assessing individual pixels, the read noise is given by the standard deviation of the signal at temporally constant illumination \cite{PTC_Li}, and the noise is dominated by shot noise when the read noise is subtracted or becomes negligibly small in relation to the other noise contributions, and before the sensor approaches saturation.

\begin{figure}[H]
\includegraphics[width=6 cm]{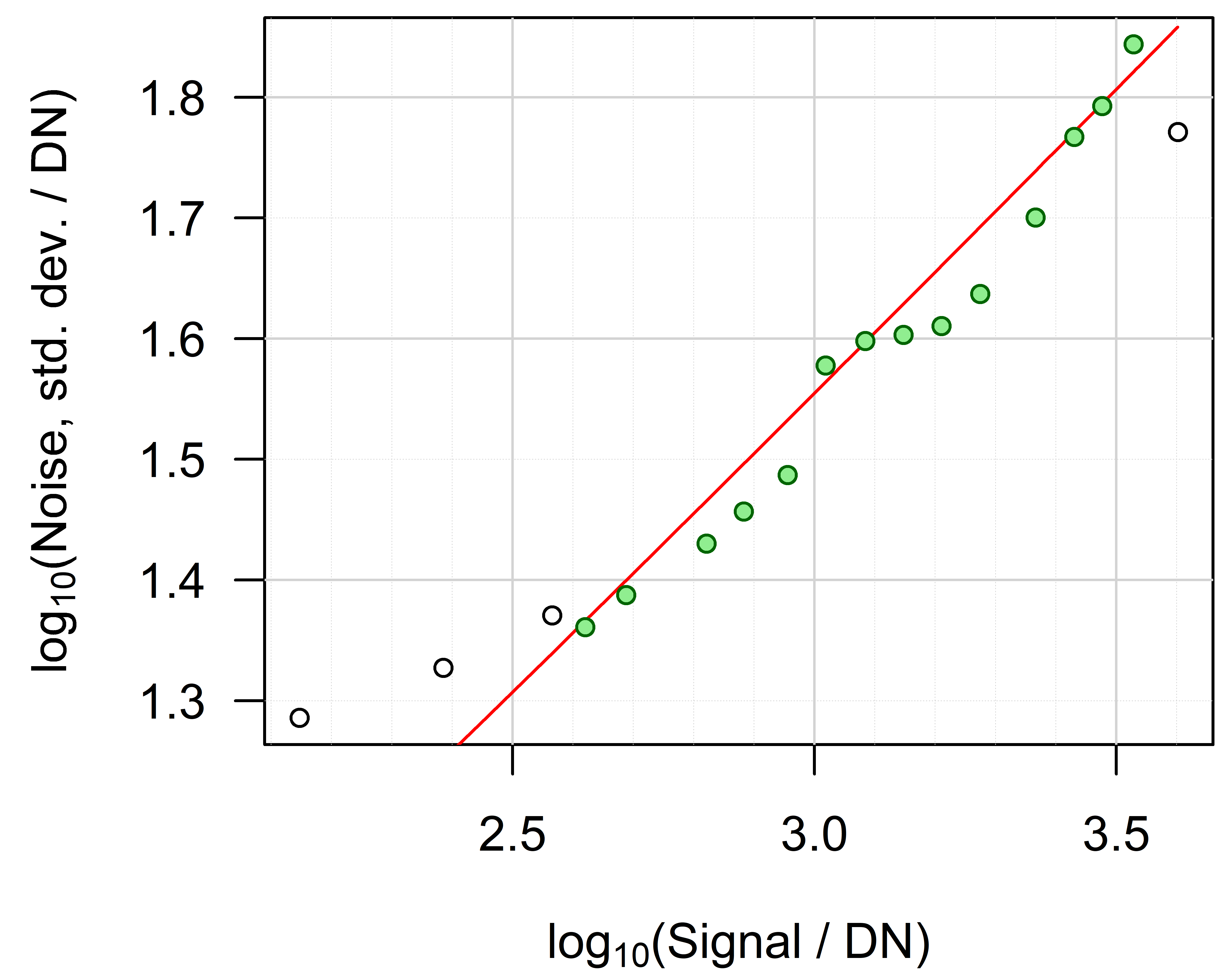}
\caption{Photon transfer curve measured by binning the camera characterization data (see text, below). From Eqn.~(\ref{eq:shot_noise}), $\log_{10}(\sigma_\textrm{shot}) = 0.5\log_{10}(S)$, i.e., in the shot-noise limited region, the logarithm of the noise is expected to scale linearly with the logarithm of the signal, with a gradient of 0.5. The data points in this region, marked in green, were fitted with a linear model, yielding a gradient of 0.51. The intercept yields an expected gain of 0.96~DN/e$^-$. The data point with the lowest signal amplitude is omitted from the graph.}
\label{fig:PTC}
\end{figure}

The photon transfer curve is a log-log plot of the noise, measured as the standard deviation of the signal, versus the signal. The region in which the total noise is dominated by shot noise scales linearly with the square root of the signal, and can be used to determine the proportionality constant between the squared shot noise (variance of the signal) and the signal, yielding the conversion factor between digital numbers and photoelectrons (see Eqn.~(\ref{eq:DNtoelectron})). To obtain data pairs of signal and variance, time series of pixels are acquired at an ideally time-constant illumination. Here, a white screen (IPS LED) of a laptop plugged into the mains was used to illuminate the camera sensor, with a piece of paper inserted between the screen and the sensor to achieve different levels of illumination. For the dark image, the camera's dust cap was screwed onto the camera. Time series of 10000 frames were used. The resulting time-variance vs. time-mean values of each pixel at the various illumination intensities were collectively binned by mean value, yielding a data point for each movie file used, but correcting for spatial inhomogeneities in the illumination. The resulting photon transfer curve is shown in Fig.~\ref{fig:PTC}.

\begin{figure}[H]
\begin{subfigure}[b]{.42\textwidth}
\caption{}
\includegraphics[height=4.7cm]{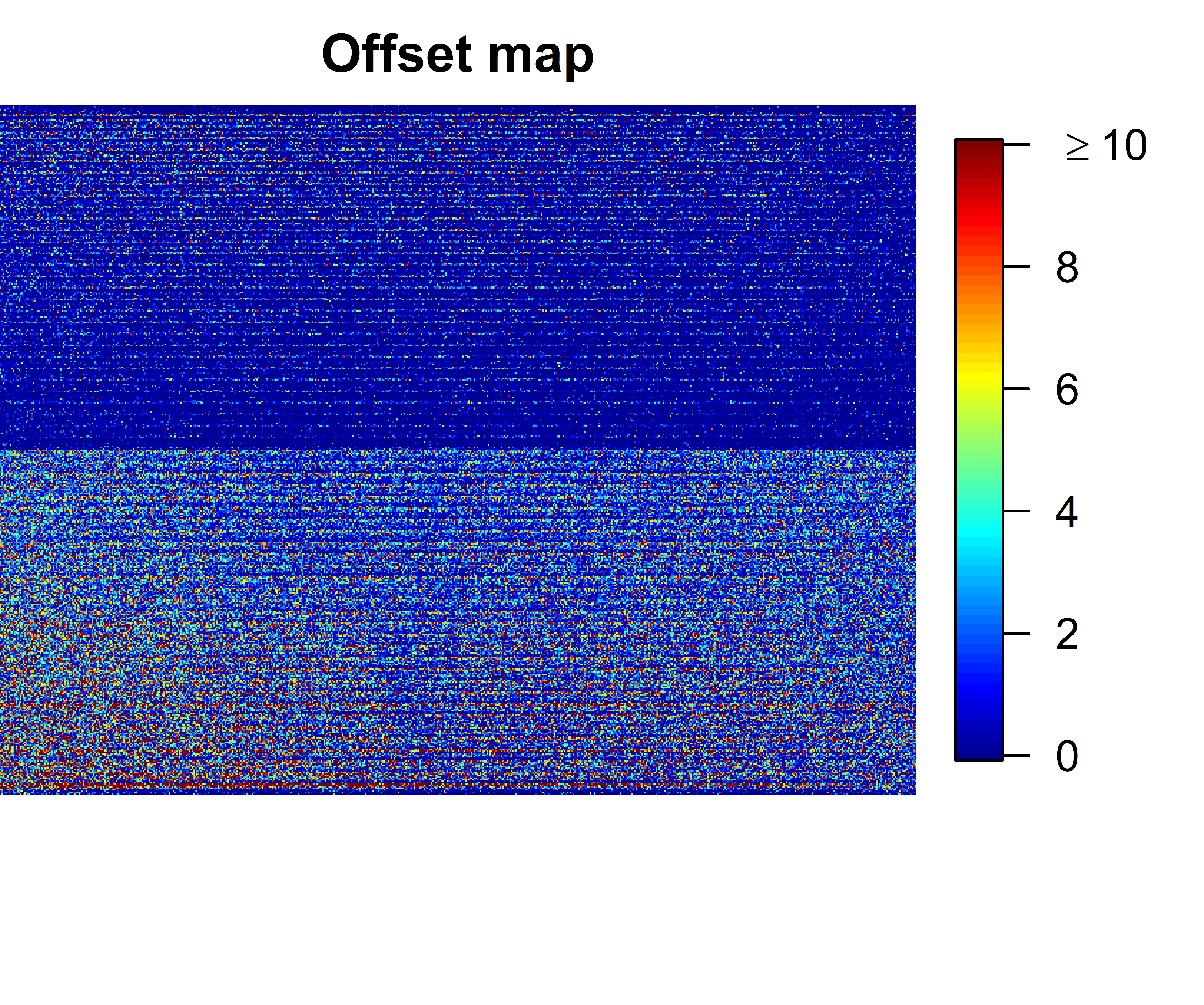}
\end{subfigure}
\begin{subfigure}[b]{.04\textwidth}
\includegraphics[height=4.7cm,trim={0 0 11cm 0},clip]{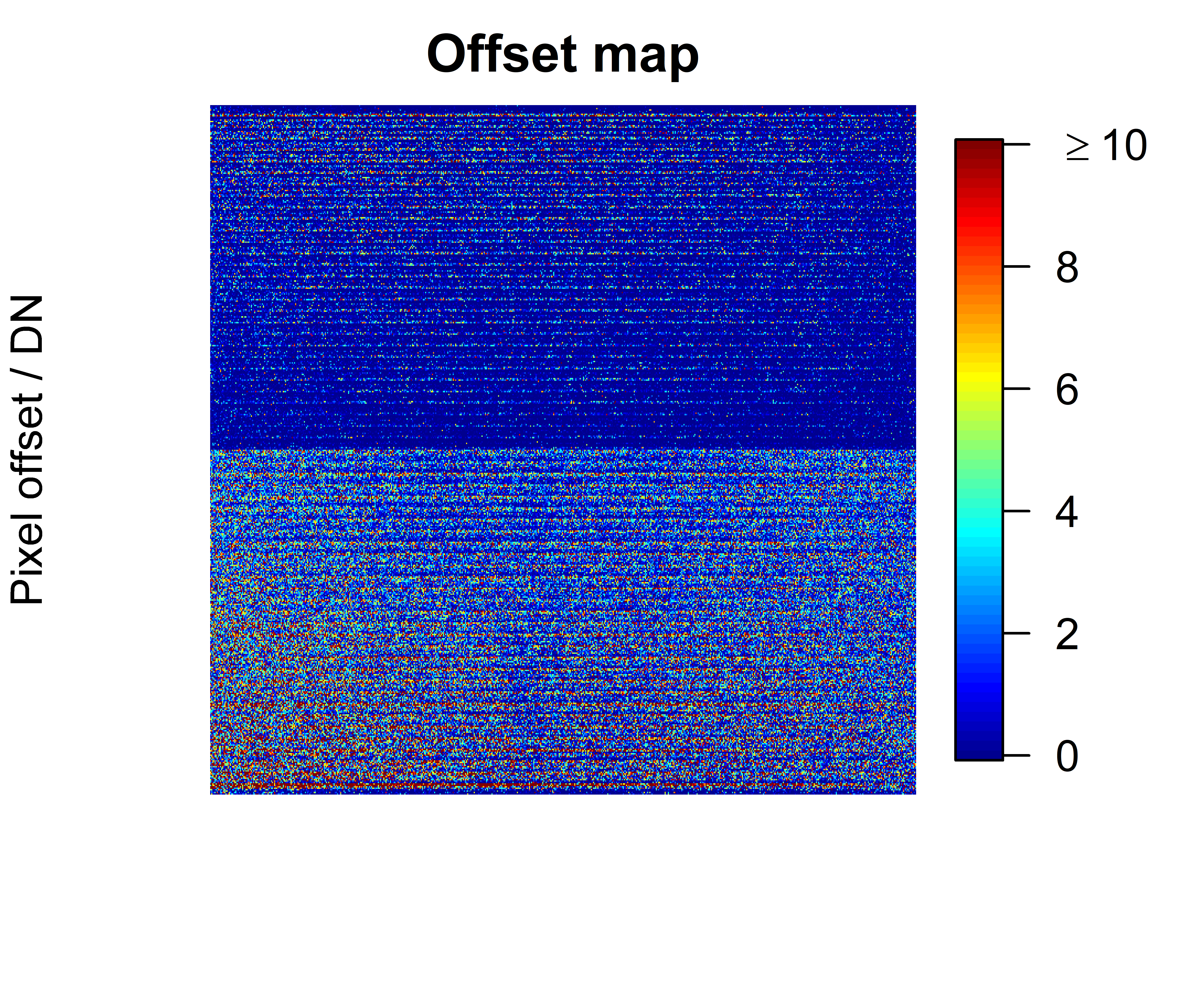}
\end{subfigure}
\begin{subfigure}[b]{.4\textwidth}
\includegraphics[height=4.7cm]{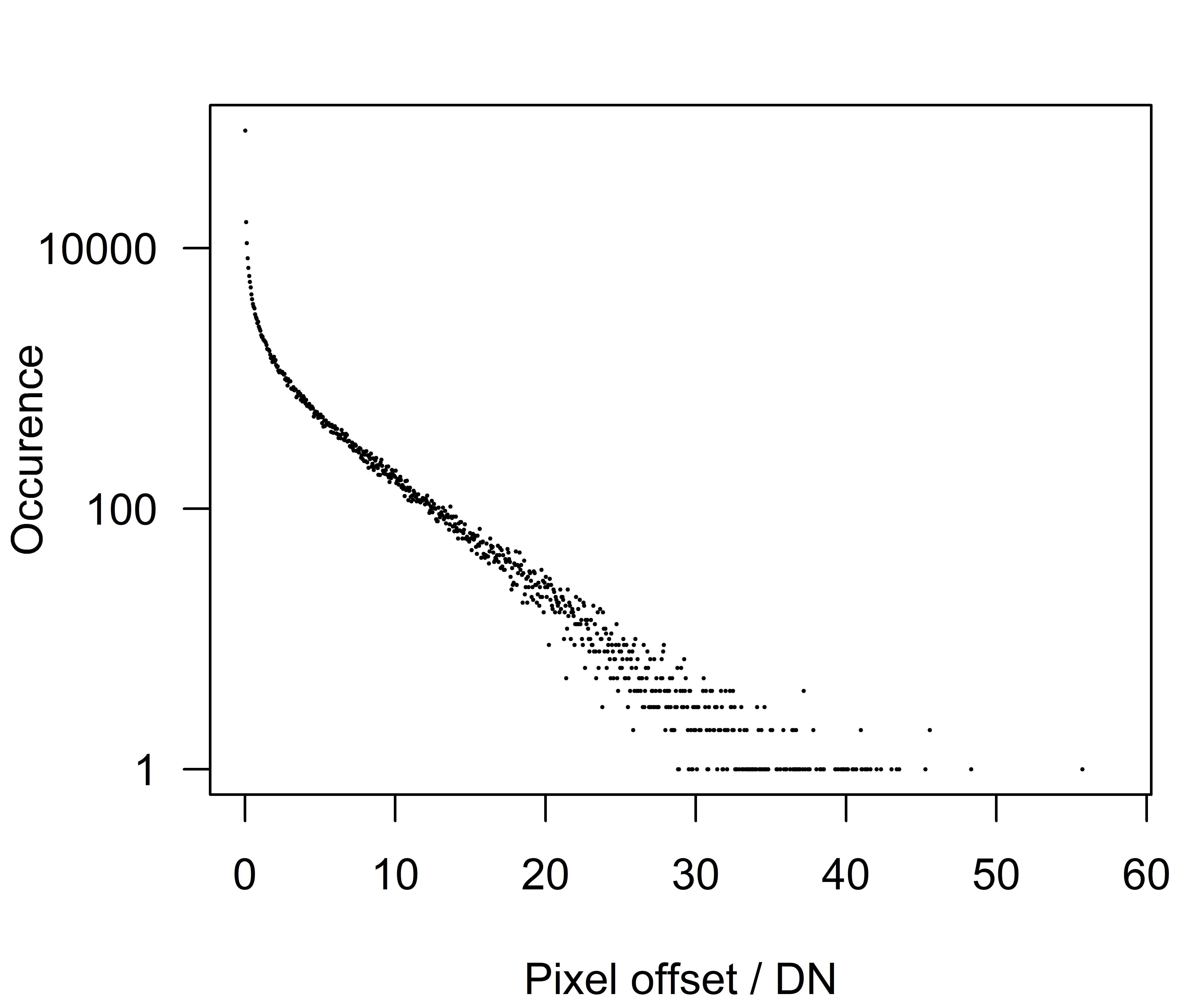}
\end{subfigure}\\
\begin{subfigure}[b]{.42\textwidth}
\caption{}
\includegraphics[height=4.7cm]{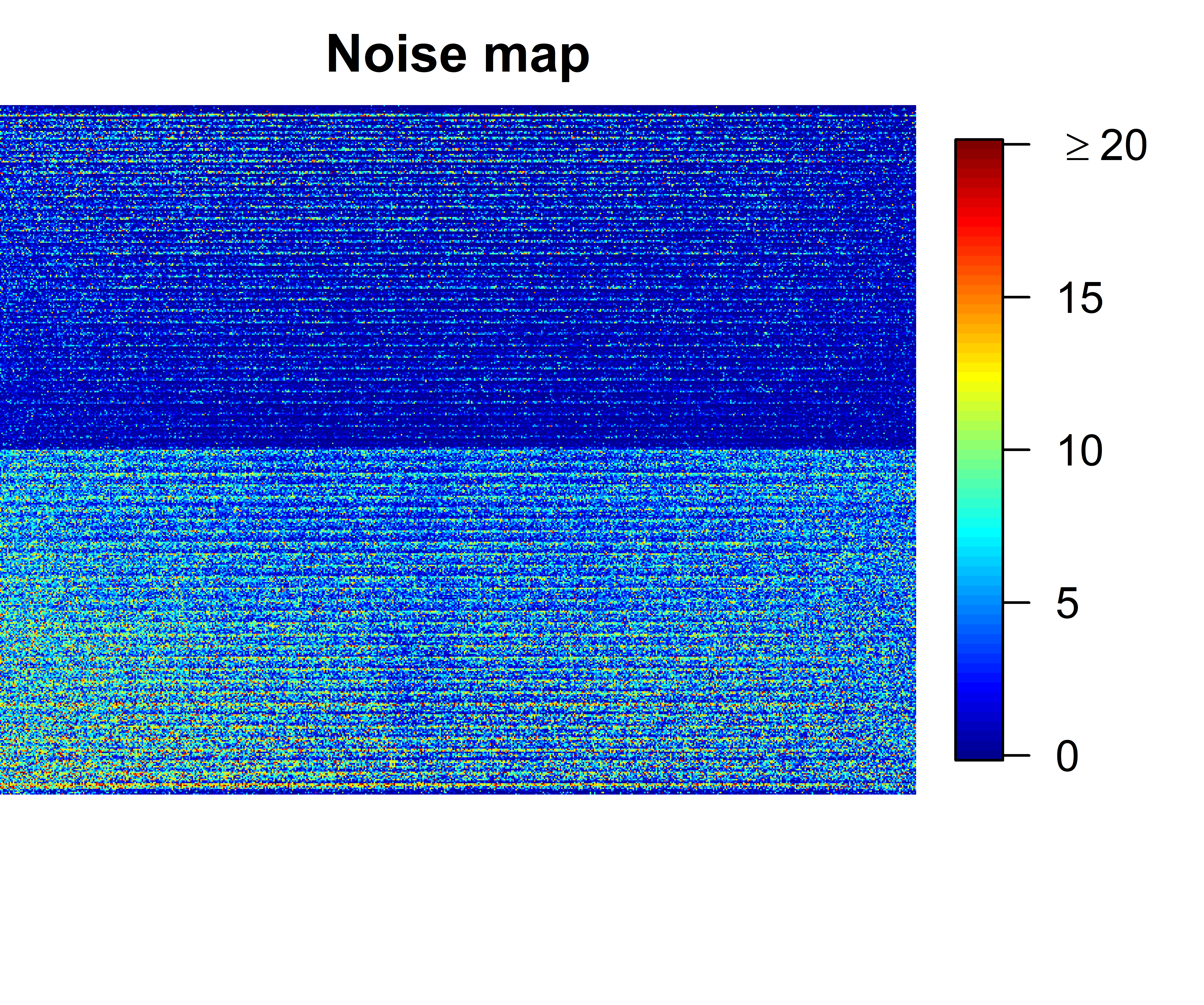}
\end{subfigure}
\begin{subfigure}[b]{.04\textwidth}
\includegraphics[height=4.7cm,trim={0 0 11cm 0},clip]{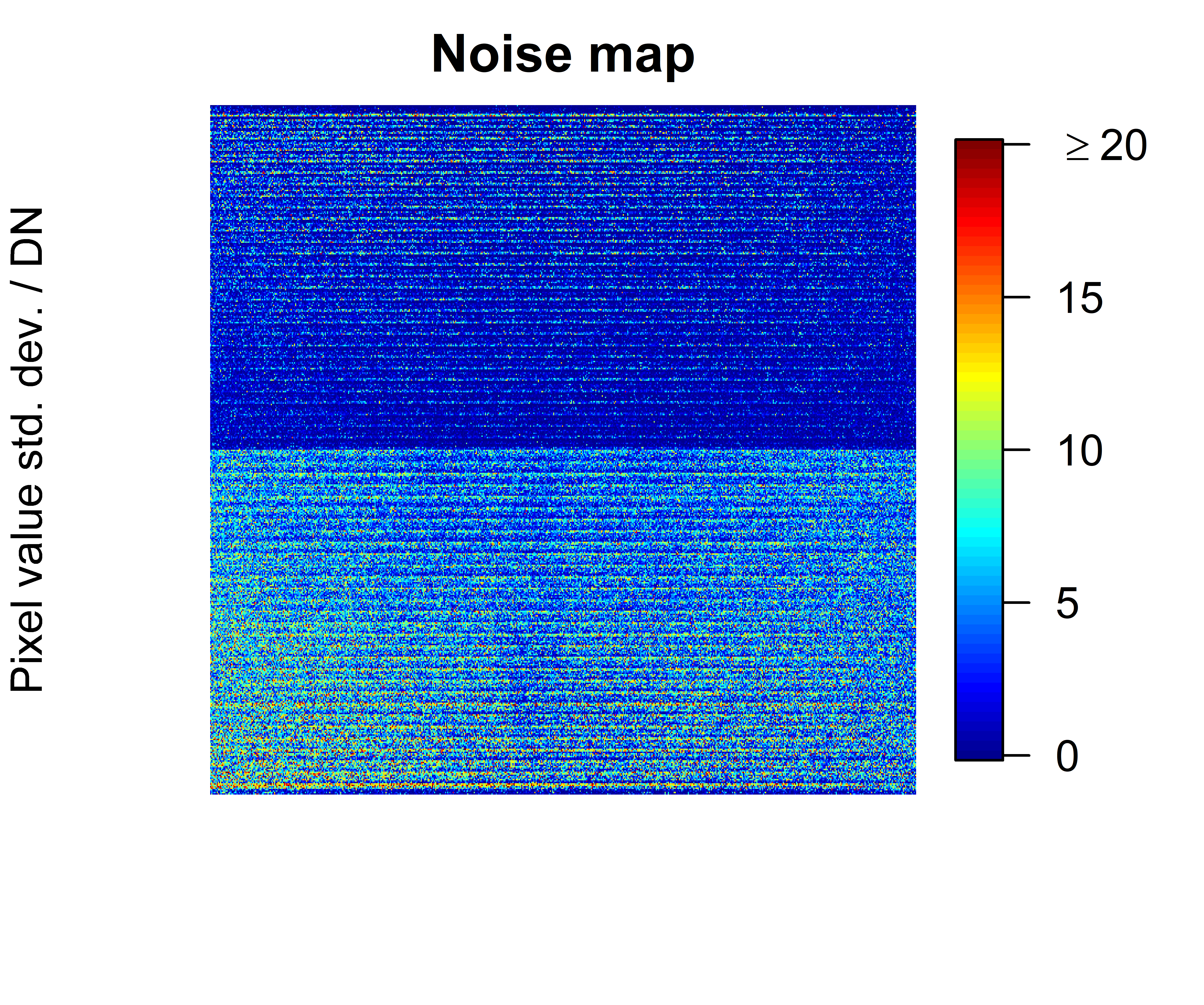}
\end{subfigure}
\begin{subfigure}[b]{.4\textwidth}
\includegraphics[height=4.7cm]{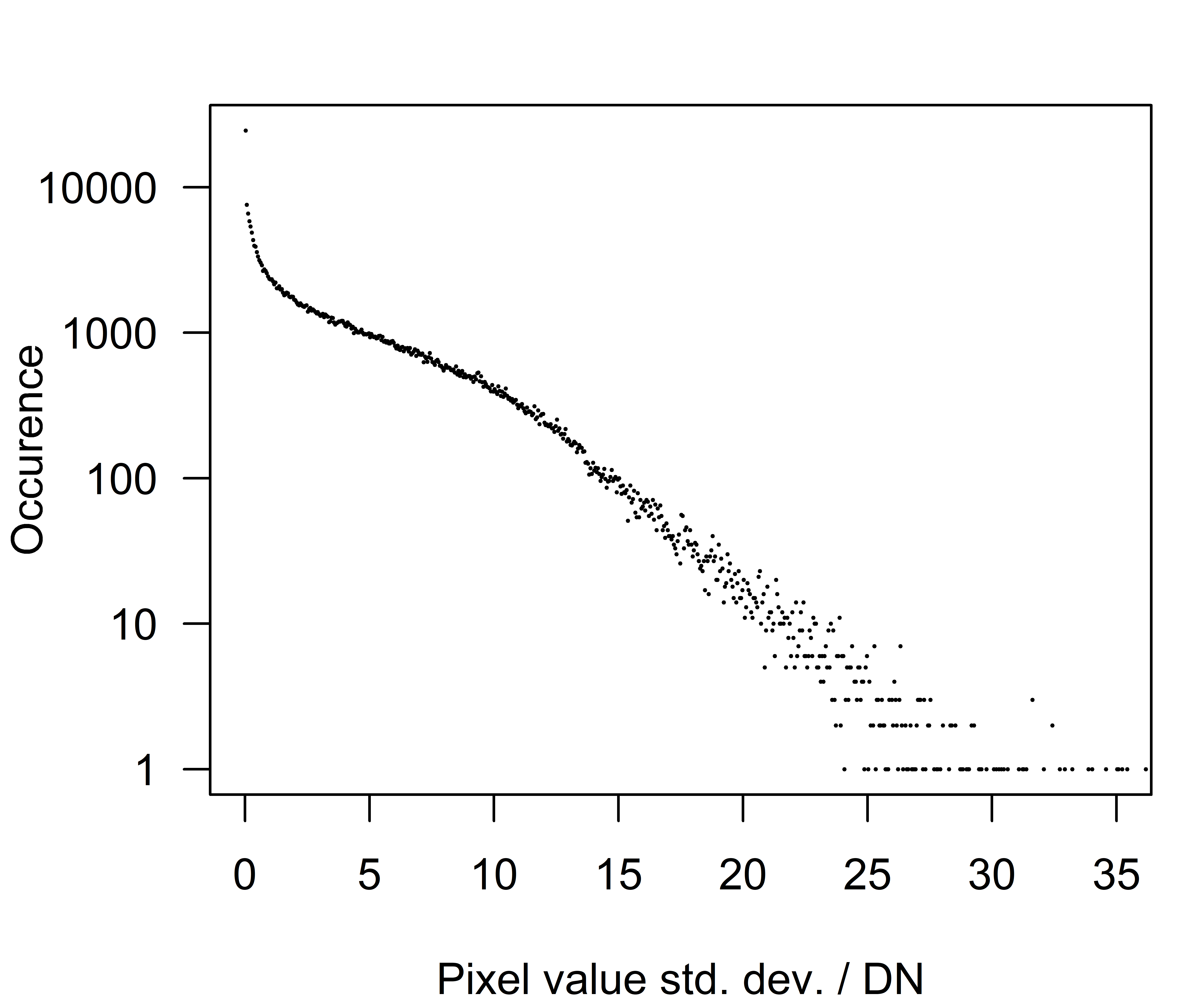}
\end{subfigure}\\
\begin{subfigure}[b]{.42\textwidth}
\caption{}
\includegraphics[height=4.7cm]{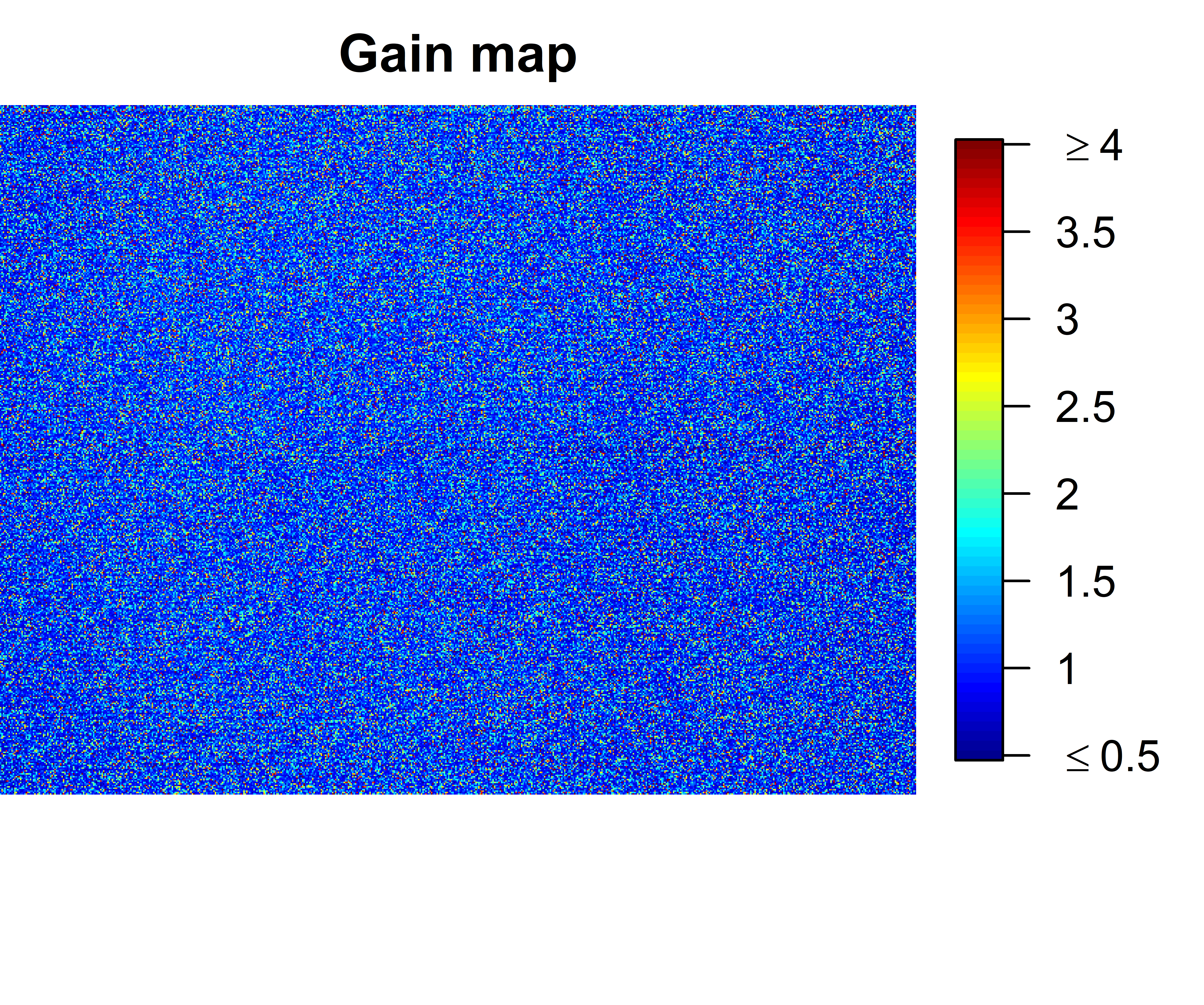}
\end{subfigure}
\begin{subfigure}[b]{.04\textwidth}
\includegraphics[height=4.7cm,trim={0 0 11cm 0},clip]{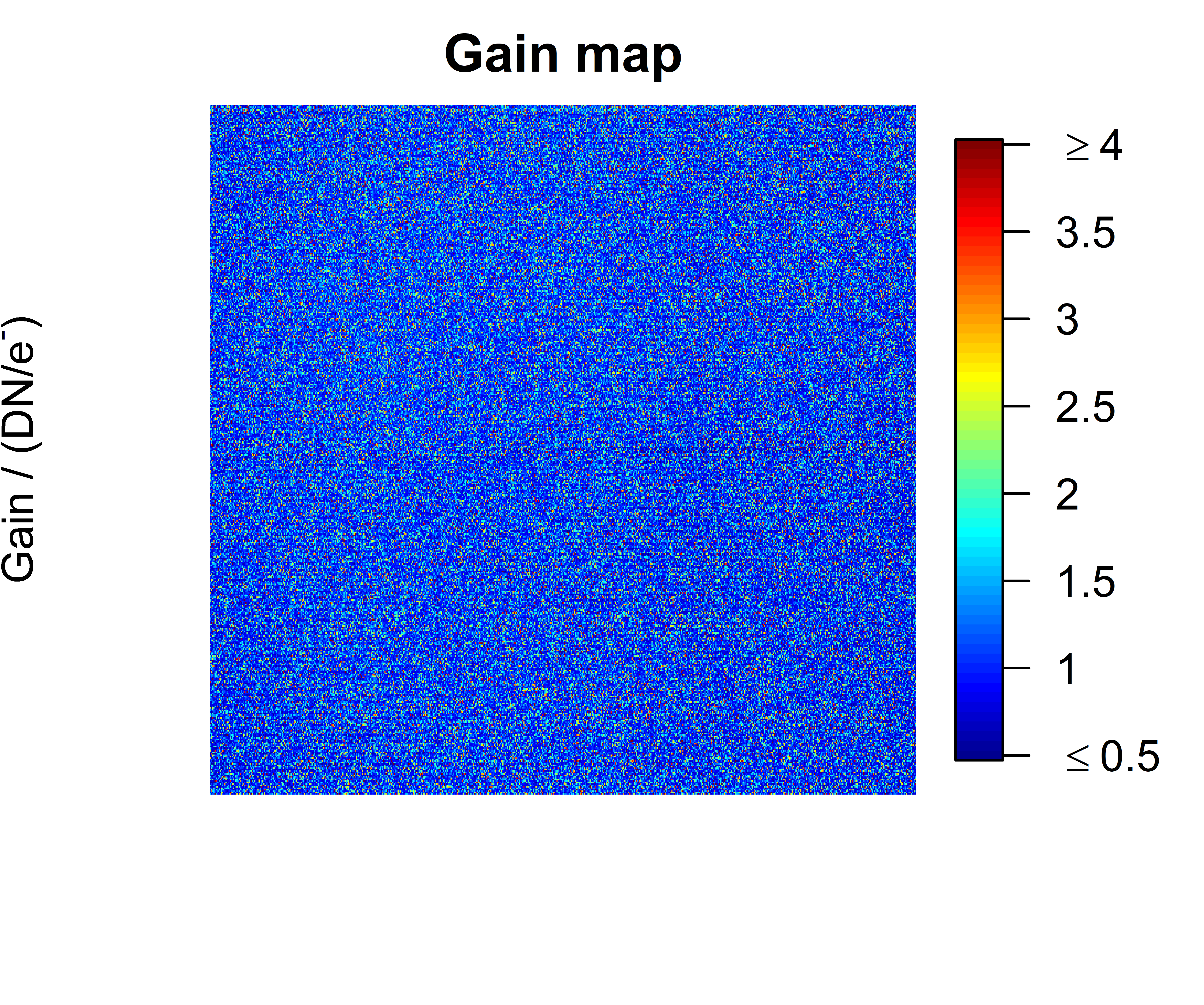}
\end{subfigure}
\begin{subfigure}[b]{.4\textwidth}
\caption{}
\includegraphics[height=4.7cm]{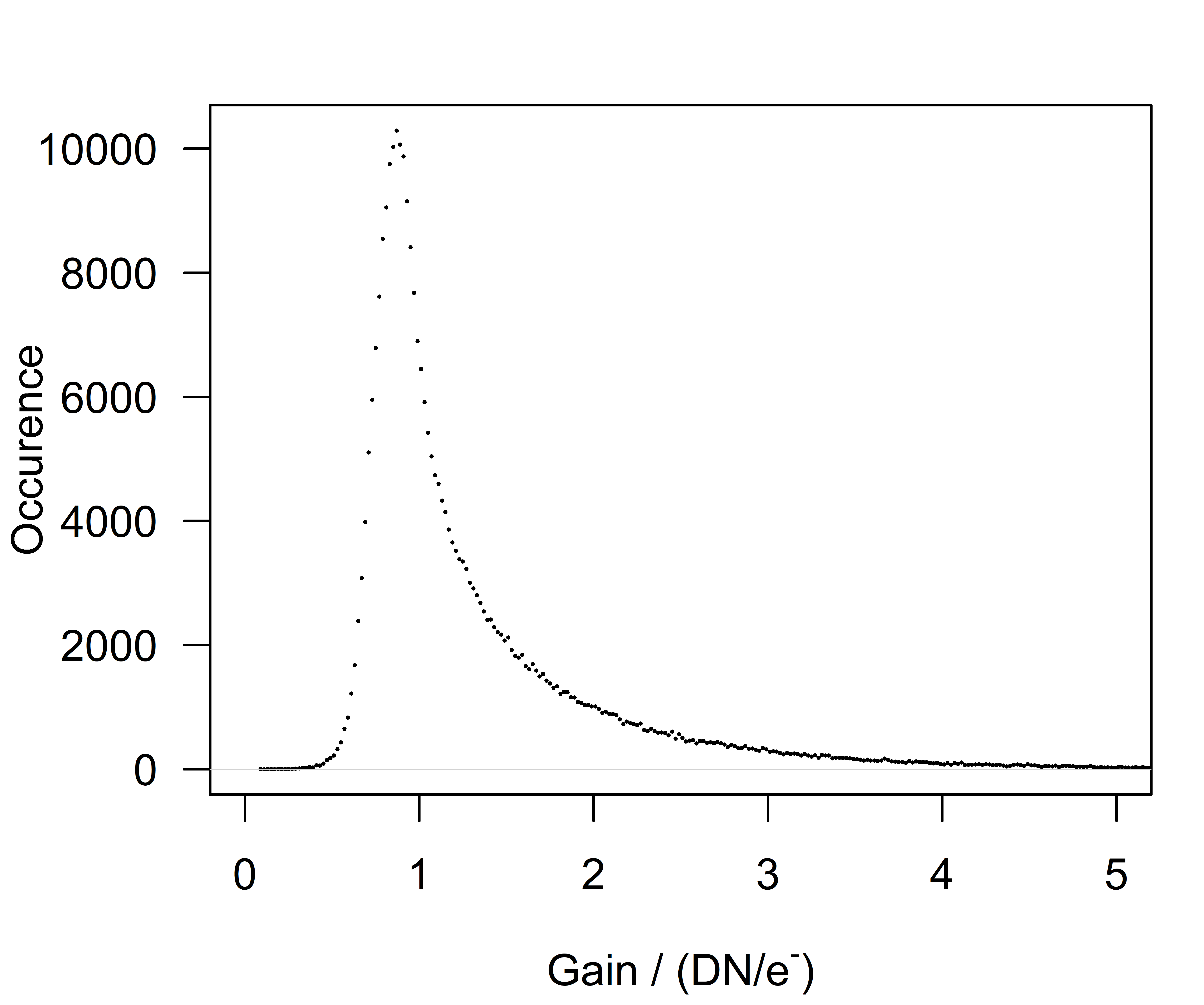}
\end{subfigure}
\caption{\textbf{Maps of the 640x480 pixel area in the centre of the camera chip used for the measurements.} a)~Dark offset, b)~dark noise, c)~gain; pixel maps (left) and corresponding histograms (right). The width of the gain distribution is likely to be overestimated because inhomogeneities in the illumination profile are likely to have an effect given the small signal window of the photon transfer curve (PTC) used, i.e. not every pixel is evaluated in the same region of the PTC. In the setup shown in Fig.~\ref{fig:TIR-DF}, this area of the camera chip corresponds to a 48~$\mu$m$\times$36~$\mu$m field of view.}
\label{fig:Fastcam_charac}
\end{figure}

To obtain a spatial map of the conversion factor, the pixels were also assessed individually (see Ref.~\cite{CMOSchar}). To calculate the gain, the offset, the dark mean pixel value, is subtracted from the rest of the data set. Offset and read noise (dark noise) and gain maps are shown in Figs.~\ref{fig:Fastcam_charac}a and b, respectively, and a gain map is shown in Fig.~\ref{fig:Fastcam_charac}c, using only the datasets in the shot noise-limited region of the photon transfer curve as identified in Fig.~\ref{fig:PTC}. Spatially averaged values can be found in Tab.~\ref{tab:Fastcam_charac}. The spatially averaged gain value is used for the entire camera chip since the distribution is quite narrow.

\begin{table}[H]
\caption{Spatially averaged (median) values of the camera characterization data.}
\begin{tabular}{l*{6}{c}r}
              			& Digital number	& Photoelectrons \\
\hline
Offset		 		& 0.59		& 0.57 \\
Noise (std. dev.)		& 2.51		& 2.41 \\
Gain				& 1.04		& 1 \\
Conversion factor $C'$	& 1			& 0.96 \\
\end{tabular}
\label{tab:Fastcam_charac}
\end{table}

The noise measured experimentally overestimates the camera noise, with temporal fluctuations of the illumination intensity and possibly other factors contributing to the measured noise. The laptop screen brightness fluctuates measurably, however, the amplitude is small ($<$1\%) relative to the amplitude of the signal.

\subsection{Materials}
\label{sec:materials}

\textbf{Glass coverslips} (Menzel, 24x50 mm, thickness 1.5) are rinsed alternatingly with Milli-Q water and ethanol (Sigma-Aldrich, \textsc{gc} grade); five times in total, then sonicated for 15 minutes in a 1:1 mixture of Milli-Q water and isopropanol (Sigma-Aldrich, \textsc{gc} grade), and before use rinsed again with Milli-Q water and dried with a jet of nitrogen gas. If they were stored between sonication and use, they were left in the solution they were sonicated in for no longer than 12 hours.

\textbf{Gold nanoparticles} with \textsc{pvp} functionalization were purchased from nanoComposix (gold nanospheres, \textsc{pvp}, NanoXact, 0.05~mg/mL) and diluted with motility buffer (see below for details), Milli-Q water or a \DDM{} solution in ratios of typically 1:50 for 80~nm particles (lot number ECP1384), 1:200 for 40~nm particles (lot number ECP1036) and 1:1500 for 20~nm particles (lot number ECP1386). The different dilutions resulted in similar concentrations for 20 and 40~nm particles; the concentration of 80~nm particles was lower by a factor of 2.

\begin{table}[H]
\caption{Specifications of stock solutions of \textsc{pvp}-functionalized gold nanoparticles (from nanoComposix): Diameter $d$ (\textsc{tem}), concentration $c$ in particles per milliliter, hydrodynamic diameter $d_\textrm H$, zeta potential $\zeta$.}
\begin{tabular}{l*{6}{c}r}
Particle type	& $d$ / nm		& $c$				& $d_\textrm H$ / nm	& $\zeta$ / mV	\\
\hline
\noalign{\vskip 0.1cm}
20~nm Au	& 19.1$\pm$2.4	& 7.4$\cdot10^{11}$	& 44					& -28			\\
40~nm Au	& 40.8$\pm$4.5	& 7.5$\cdot10^{10}$	& 68.7				& -31			\\
80~nm Au	& 83$\pm$10	& 9.0$\cdot10^9$		& 105					& -17			\\
\end{tabular}
\label{tab:AuNPspecs}
\end{table}

\textbf{Solutions of particles at surfactant-free (MQ) interfaces} are diluted in 18.2~M$\Omega$~cm Milli-Q water; solutions of particles intended \textbf{for observation on glass} are diluted with a buffer (\textsc{mb}) as particles in water do not attach to the glass surface. \textsc{mb} is prepared by mixing solutions of 80~m\M \textsc{mops}, 20~m\M MgCl$_{2}$ and 0.4~m\M \textsc{egta}, where the \textsc{mops} and \textsc{egta} solutions are  brought to pH 7.3 by adding HCl. \textsc{mops} (ultra high purity grade), MgCl$_2$ (anhydrous, high purity grade) and \textsc{egta} (ultra high purity grade) were purchased from Amresco. \textbf{Solutions of particles at surfactant-laden (DDM) interfaces} are prepared by adding \textit{n}-dodecyl-$\beta$-\textsc{D}-maltoside (\DDM{}): The stock solution is diluted with a 1~mg/mL solution of \DDM{} in Milli-Q (typically 0.2~m\M; the critical micelle concentration is 0.17~m\M). Solutions of \DDM{} must be replaced regularly; contaminants appear to accumulate in them within days or weeks. The \DDM{} was purchased from Anatrace (sol-grade).

\textbf{Immersion oil} for the lower phase of interface samples was purchased from Cargille (refractive index 1.5230$\pm$0.005 at 589.3~nm and 25~$\degree$C, code 1160, lot number 012490). The density was 1.076~g~cm$^{-3}$, the kinematic viscosity 41~cSt and the surface tension 35~dynes/cm; all at 25~$\degree$C. It was centrifuged at a relative centrifugal force of 17200$\times$g for 30~minutes to remove contaminants. High-viscosity immersion oil used for developing the interface protocol was purchased from Olympus (\textsc{immoil}-\textsc{f}30\textsc{cc}). The density was 0.9169~g~cm$^{-3}$ (at 15~$\degree$C); the kinematic viscosity was 1.73~cSt (at 40~$\degree$C).

\textbf{Aperture grids} made of copper were obtained from Gilder Grids (\textsc{ga}1500-\textsc{c}3, aperture width 1500~$\mu$m). The manufacturer states that the thickness of the grids increases with aperture size, varying between 25$\pm$3~$\mu$m for the 75~$\mu$m aperture and 50$\pm$5~$\mu$m for the 2000~$\mu$m aperture, leading to an estimate of 44$\pm5$~$\mu$m for the 1500~$\mu$m aperture. Previously, similar \textsc{ems}~1500-\textsc{cu} grids have been used. \textsc{tem} grids were also purchased from Gilder Grids (\textsc{g}100), with an outer diameter of 3.05~mm and hole and bar widths of 205 and 45~$\mu$m, respectively. The thickness of the grids varied with the mesh repeat as for the aperture grids, with an estimated thickness of 23$\pm5$~$\mu$m for the grids I used.

\textbf{PDMS gaskets} were prepared using the Sylgard 184 silicone elastomer kit. The base and curing agent were mixed at a ratio of 10:1 and centrifuged at 2500~rpm for 15 minutes to remove air bubbles. The mixture was poured into a Petri dish and placed in an oven at 60\degree C for several hours. The desired height of the resulting sheets was approximately 1~mm. A custom-built hole puncher was used to remove cylindrical portions and a scalpel was used to cut around them, leaving pieces of silicone with a circular aperture. The finished gaskets were sonicated using the same mixture of solvents as the coverslips, then rinsed with Milli-Q water, dried with nitrogen and stored in a Petri dish with a lid until use.


\subsection{Diffusion analysis supplement}

\begin{figure}[H]
\begin{subfigure}{.43\textwidth}
\centering
\caption{}\label{fig:20MQshoulder}
\includegraphics[width=5cm]{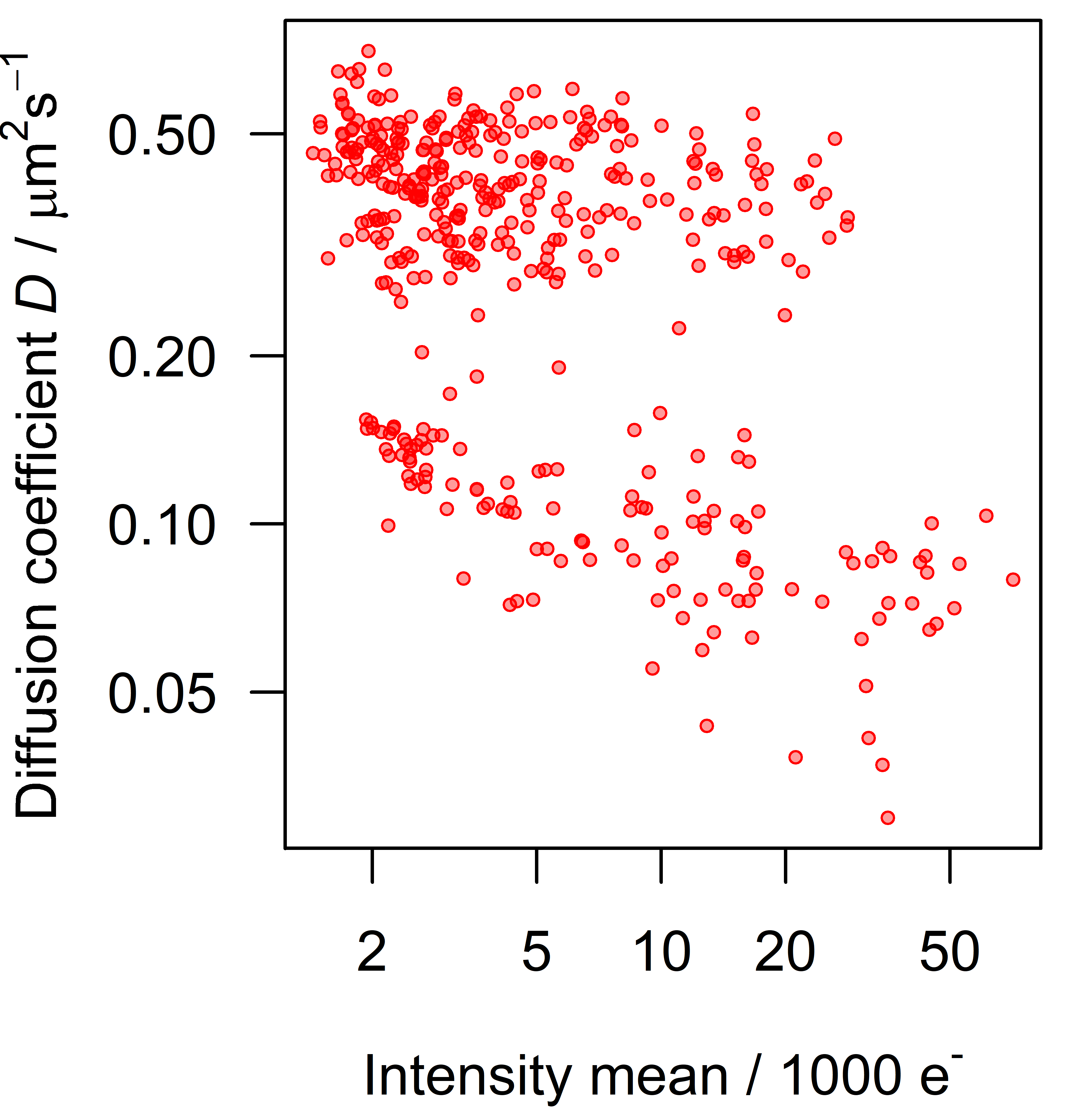}
\end{subfigure}
\begin{subfigure}{.43\textwidth}
\centering
\caption{}\label{fig:80MQshoulder}
\includegraphics[width=5cm]{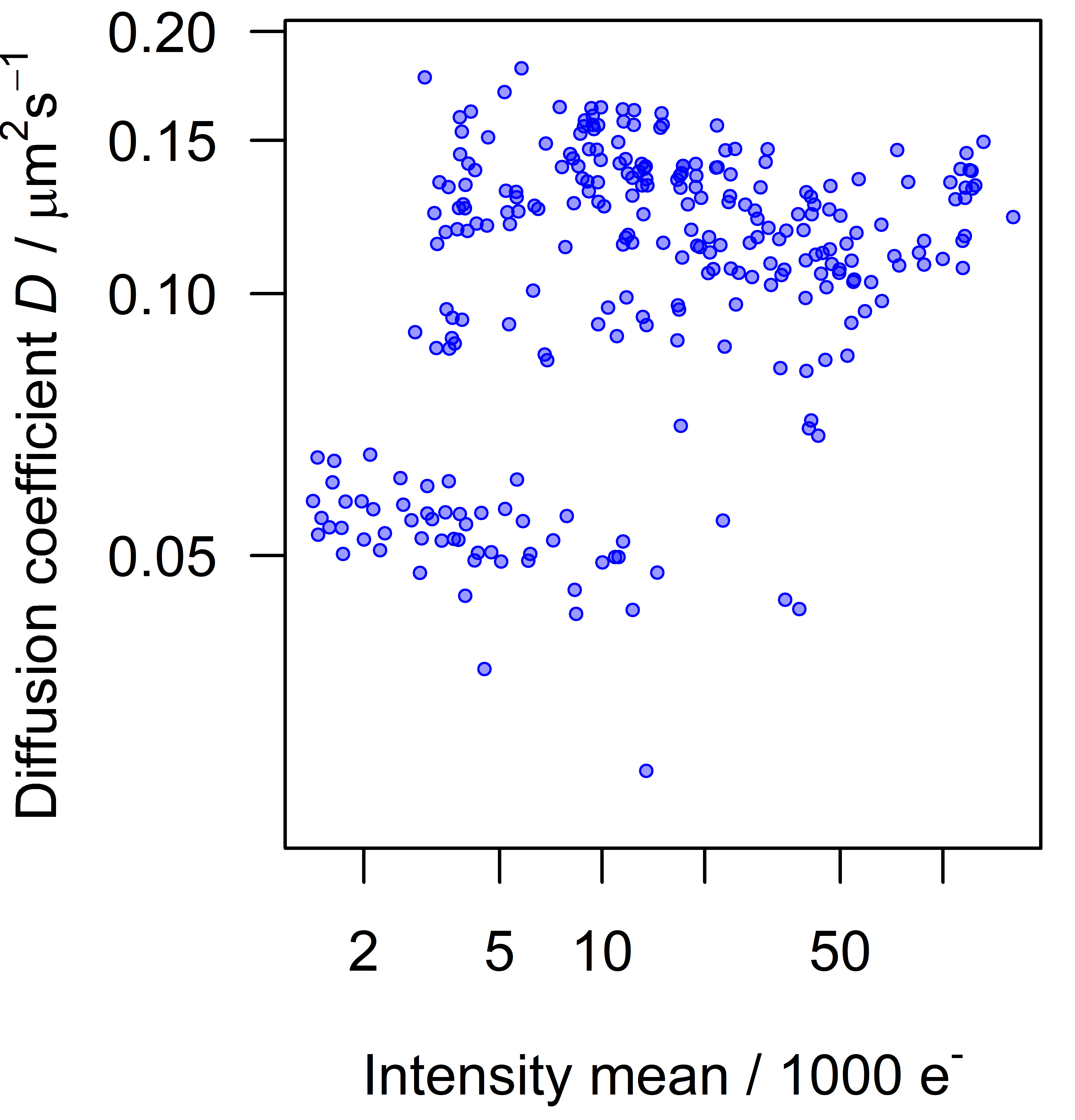}
\end{subfigure}
\caption{Diffusion coefficient vs. trajectory-averaged intensity for a) 20~nm particles and b) 80~nm particles at \MQ{} interfaces. The data set is the same as in Fig.~\ref{fig:data_D}.}
\label{fig:difcoefshoulders}
\end{figure}

\begin{figure}[H]
\begin{subfigure}{.48\textwidth}
\caption{}\label{fig:data_alphaMQ}
\includegraphics[width=5.8cm]{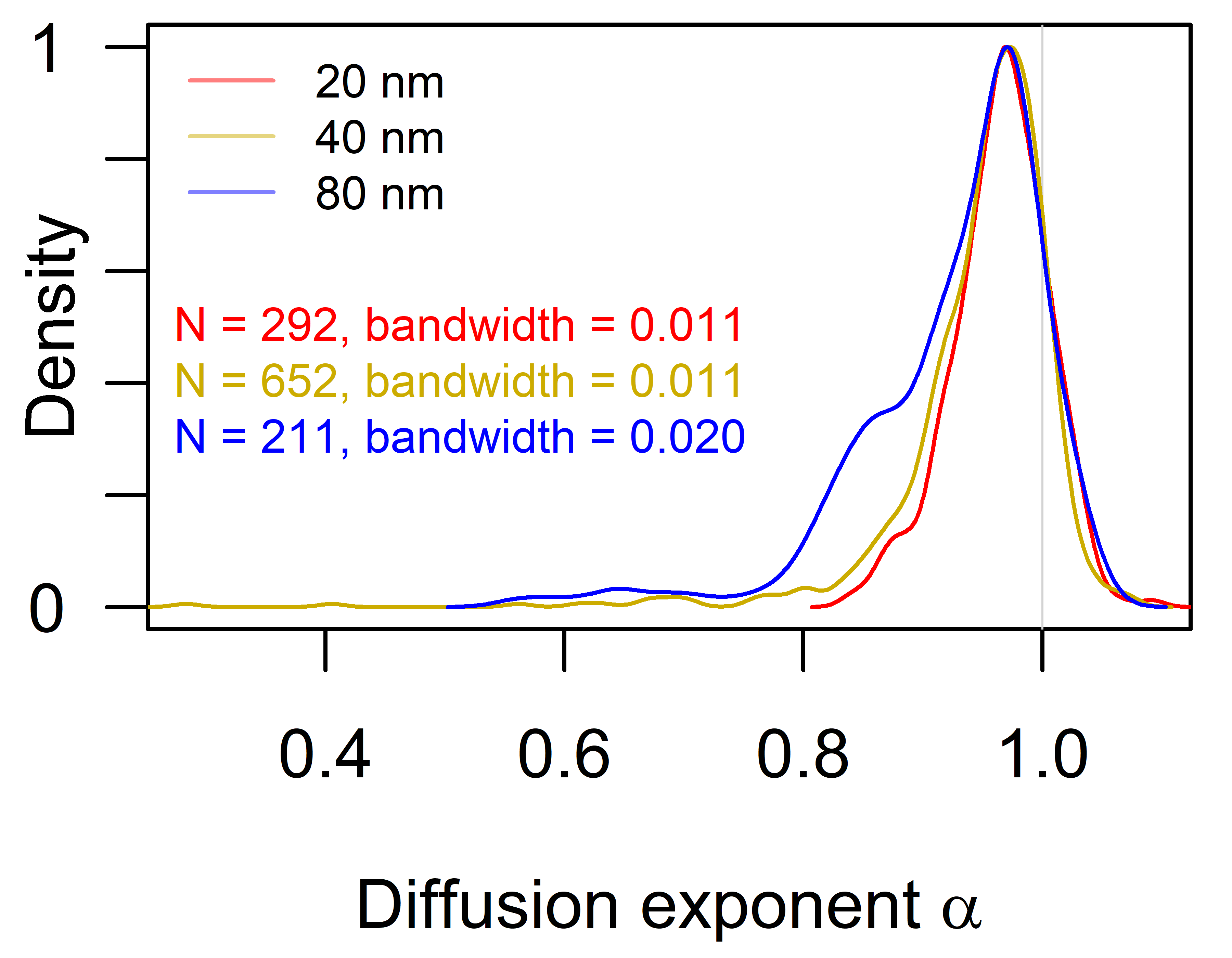}
\end{subfigure}
\begin{subfigure}{.48\textwidth}
\caption{}\label{fig:data_alphaDDM}
\includegraphics[width=5.8cm]{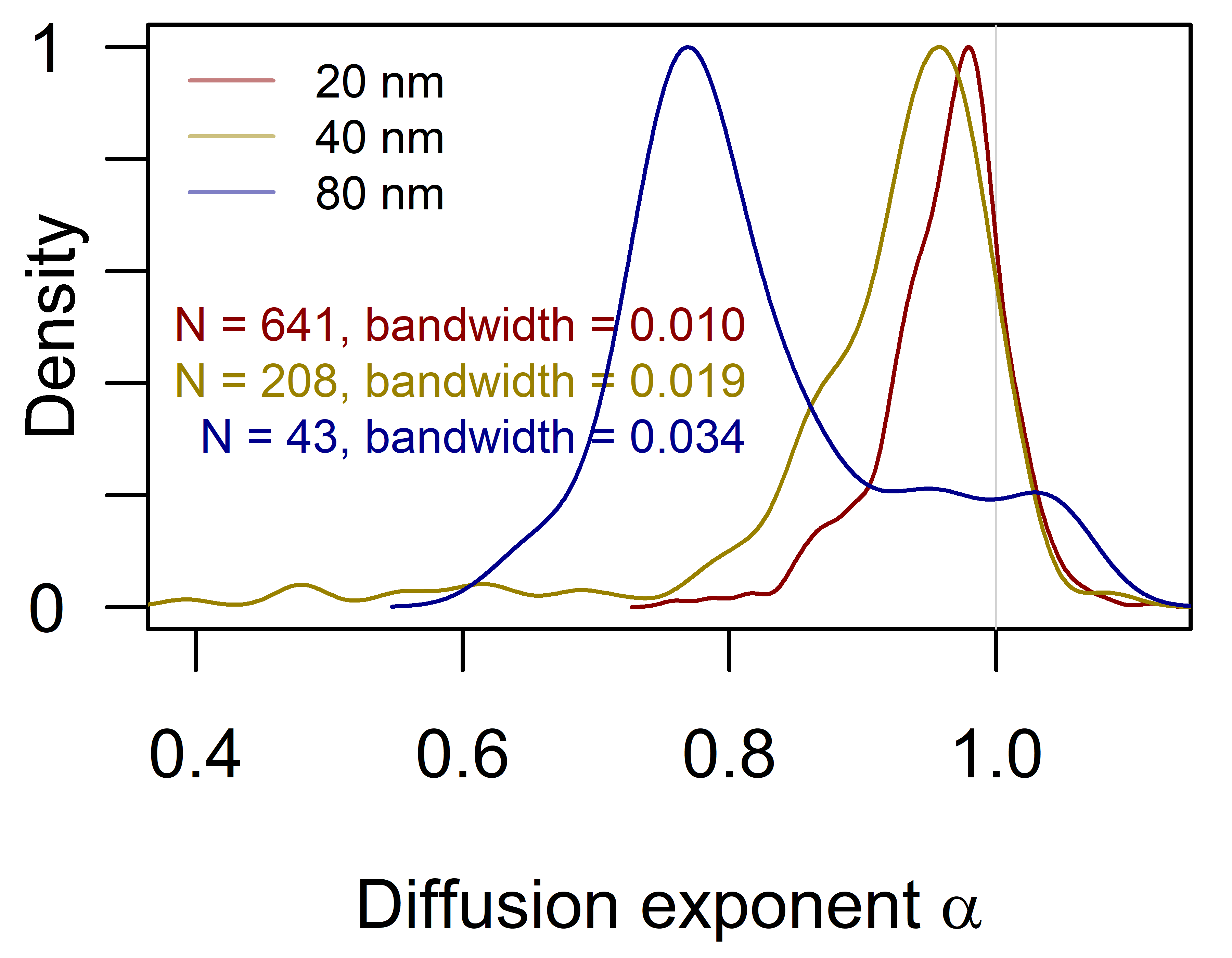}
\end{subfigure}
\begin{subfigure}{.48\textwidth}
\caption{}\label{fig:data_rogMQ}
\includegraphics[width=5.8cm]{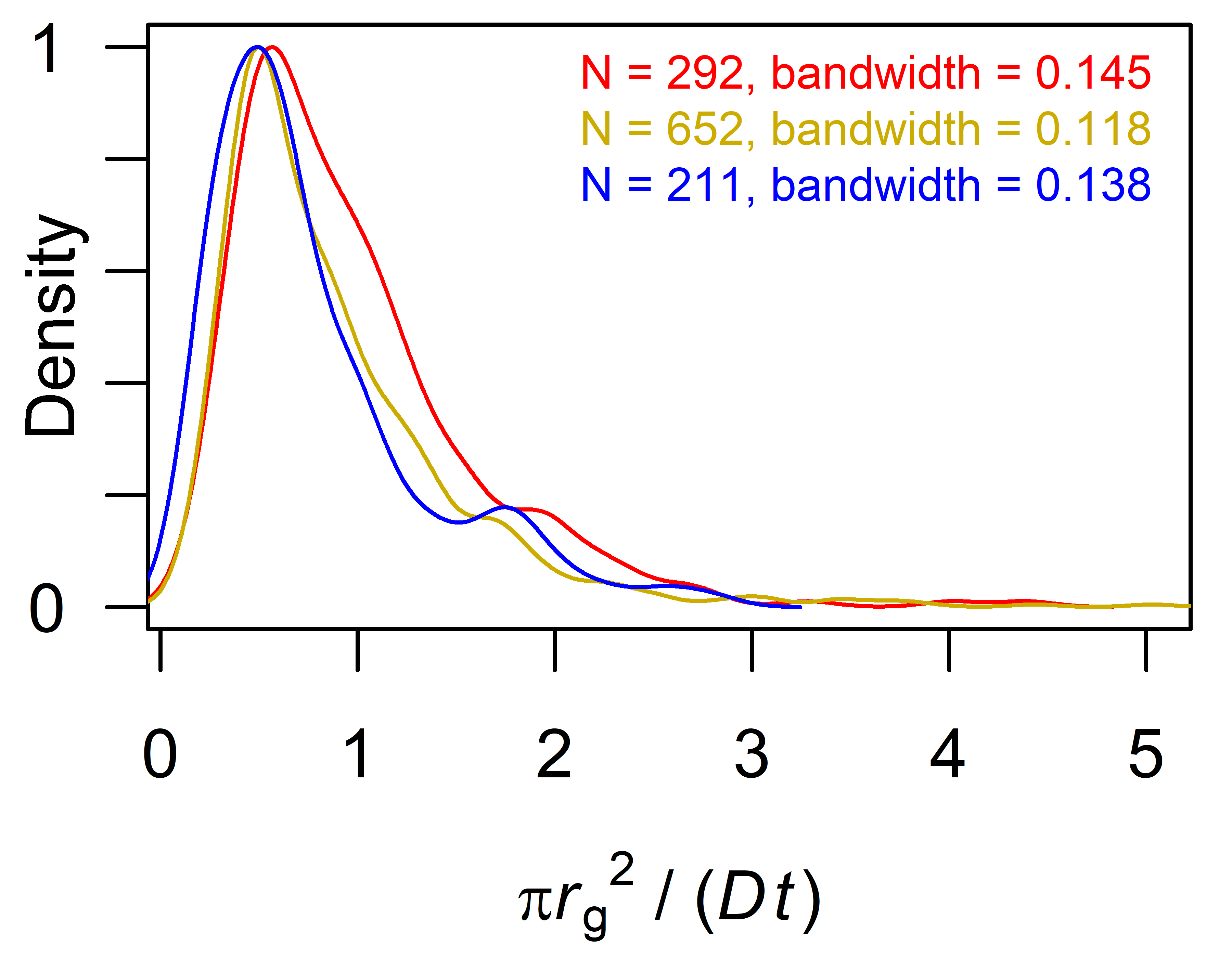}
\end{subfigure}
\begin{subfigure}{.48\textwidth}
\caption{}\label{fig:data_rogDDM}
\includegraphics[width=5.8cm]{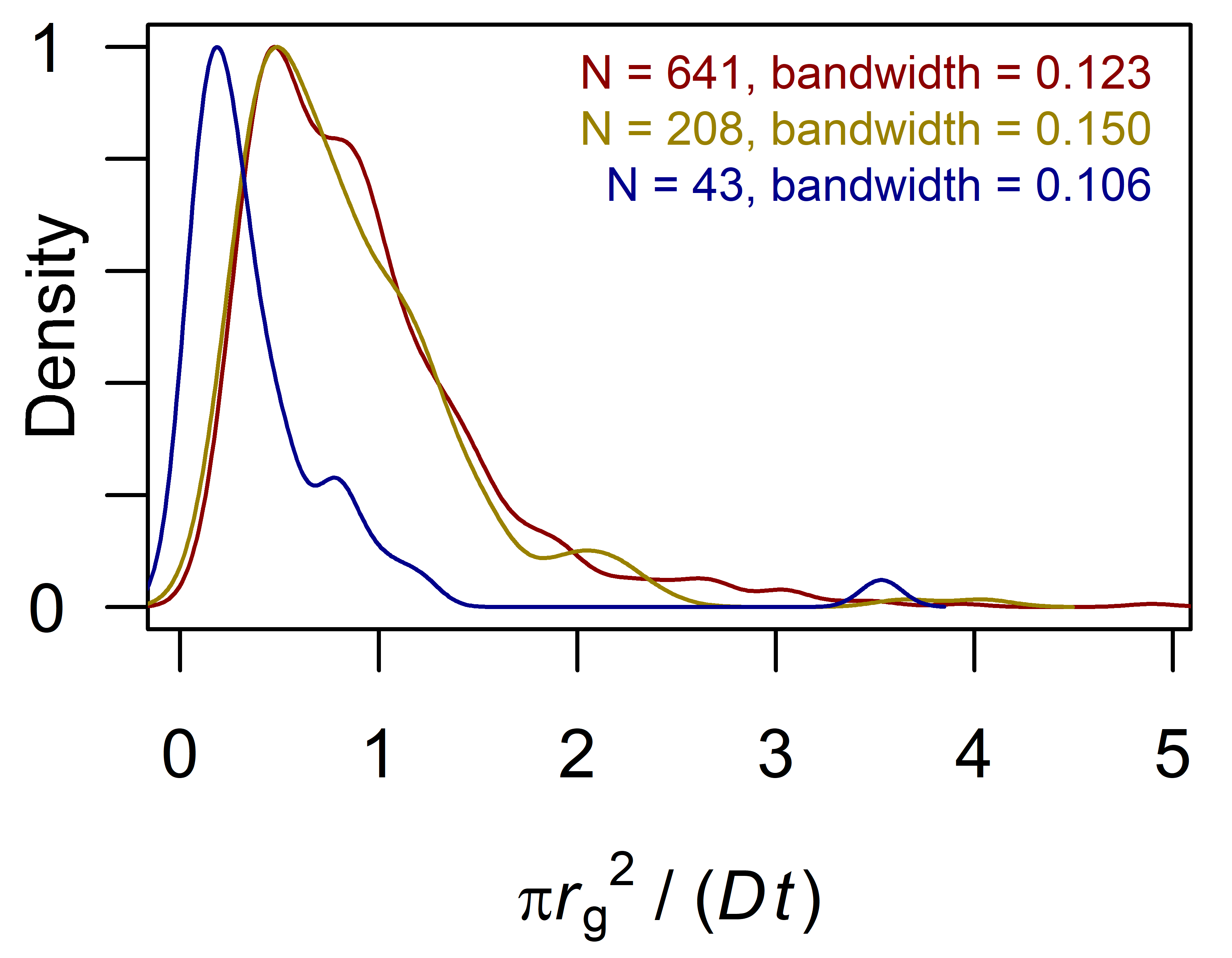}
\end{subfigure}
\begin{subfigure}{.27\textwidth}
\caption{}\label{fig:alphmean}
\includegraphics[width=3cm]{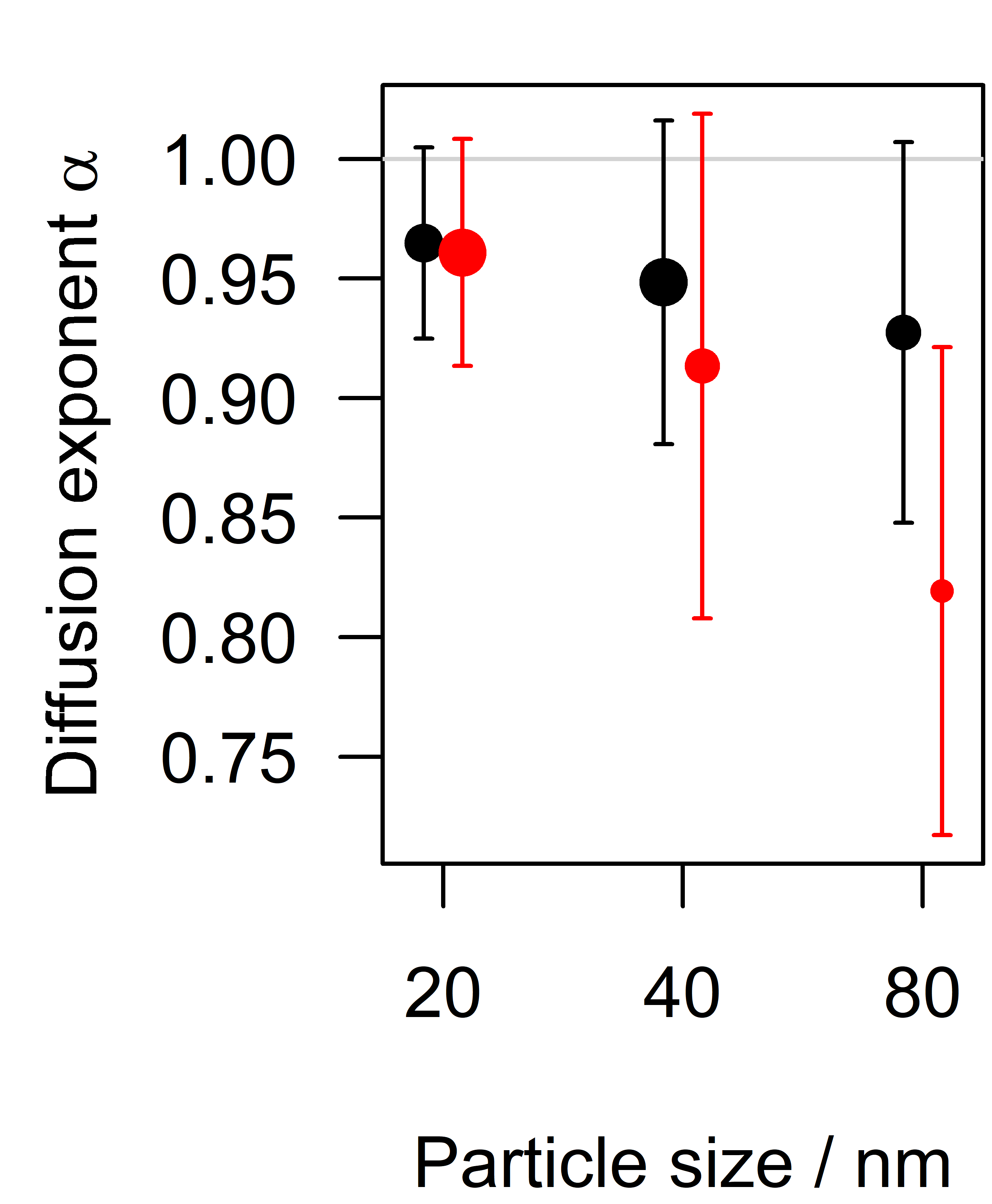}
\end{subfigure}
\begin{subfigure}{.27\textwidth}
\caption{}\label{fig:rogmean}
\includegraphics[width=3cm]{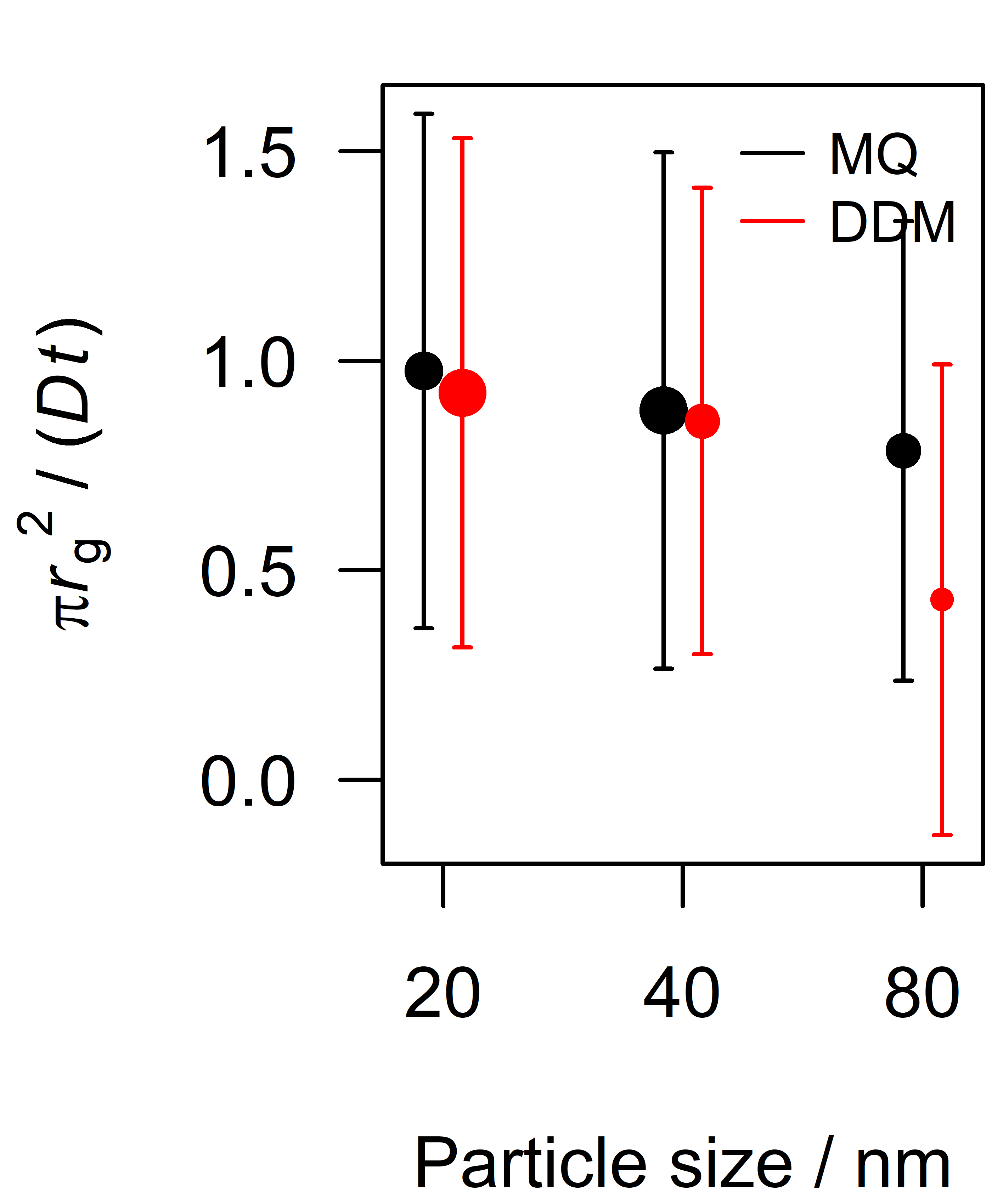}
\end{subfigure}
\caption{\textbf{Visualizing subdiffusive motion.} a)-b)~Distributions of $\alpha$ at \MQ{} (a) and \DDM{} (b) interfaces. c)-d)~Distributions of area sampled ($r_\textrm g$ is the radius of gyration: $r_\textrm g = \overline{\sqrt{(\vec{r}-\vec r_0)^2}}$) divided by diffusion coefficient and trajectory length in time, at \MQ{} (c) and \DDM{} (d) interfaces. e)-f)~Mean values of the above quantities for each sample type.}
\label{fig:data_alpha}
\end{figure}

Larger particles at \DDM{} interfaces appear to sample a smaller area compared to the length of their trajectories. This behavior can be quantified using the diffusion exponent $\alpha$: The mean squared displacement in 2-dimensional Brownian motion is given by $\textrm{\textsc{msd}} = 4D\tau^\alpha$, with $\alpha$ equal to 1. Subdiffusive motion results in $\alpha<1$. In addition, a smaller area sampled with a given duration and diffusion coefficient are also characteristics of confinement. The two quantities are shown in Fig.~\ref{fig:data_alpha}.

\begin{figure}[H]
\begin{subfigure}[b]{.48\textwidth}
\caption{}
\includegraphics[width=5.5 cm]{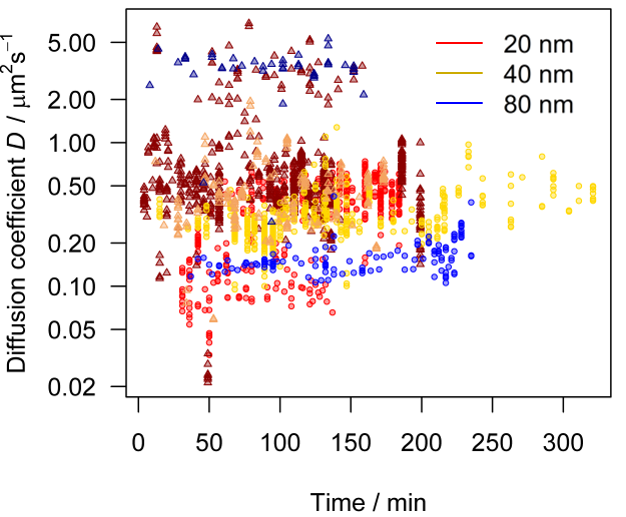}
\end{subfigure}
\begin{subfigure}[b]{.48\textwidth}
\caption{}
\includegraphics[width=5.5 cm]{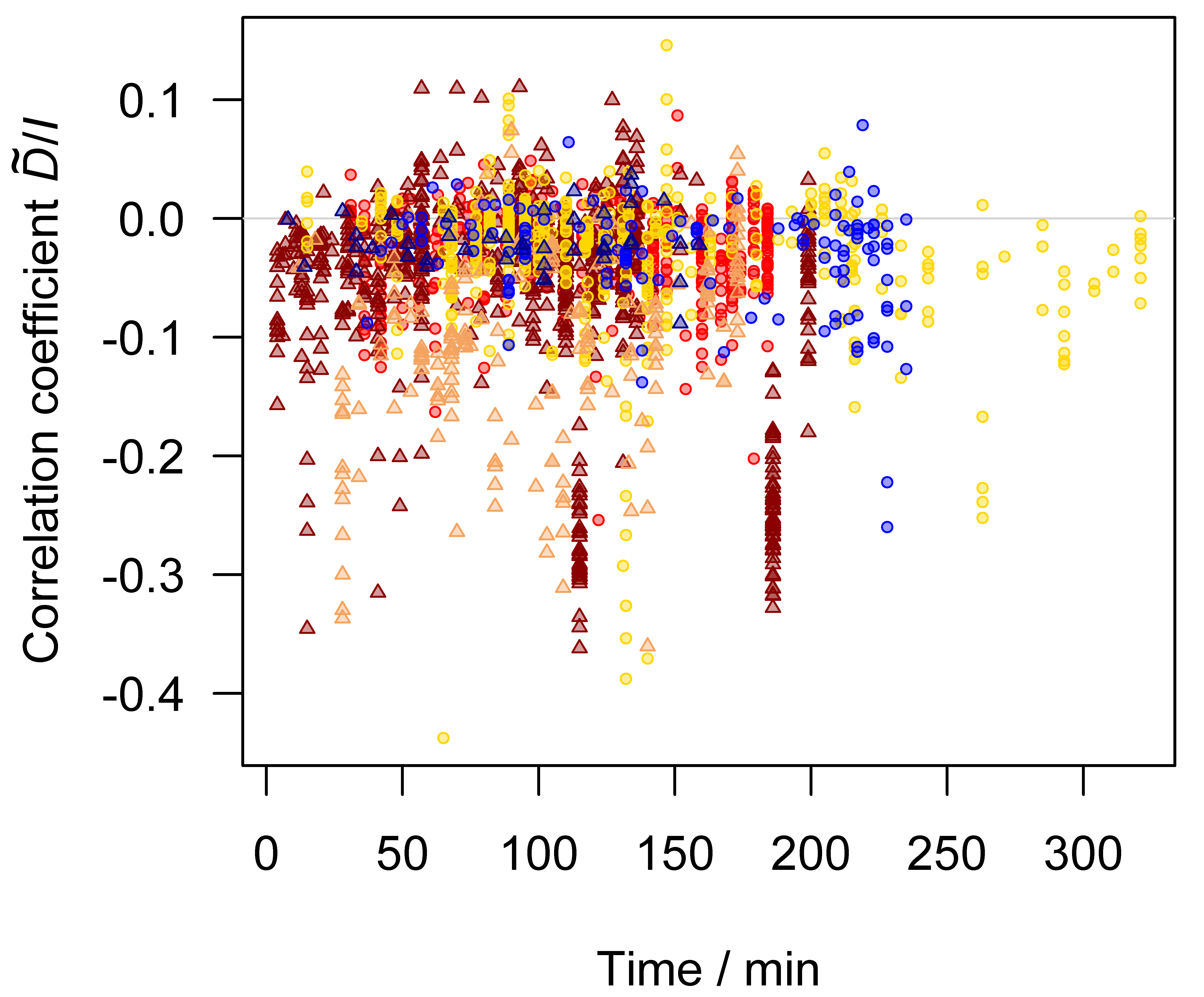}
\end{subfigure}
\caption{\textbf{Time dependence of diffusive behavior.} Diffusion coefficient (a) and correlation coefficient between intensity and instantaneous diffusion coefficient (b) for each trajectory plotted against the time elapsed between the measurement and the making of the interface. Circles represent \MQ{} interfaces and triangles \DDM{} ones; the colours in the legend correspond to \MQ{} particles and particles at \DDM{} interfaces are represented in corresponding darker colours. Three of the 22 interfaces considered in Tab.~\ref{tbl:data_agg_breakdown} are not included because the time of making the interface is not known.}
\label{fig:timeinterface}
\end{figure}

To investigate whether external conditions such as irradiation over an extended period of time or flows at the interface intensifying with evaporation (and therefore time) affect the measurements, I show aggregated data on diffusive behavior in Fig.~\ref{fig:timeinterface}.

There is minor in-plane acceleration of particles at \MQ{} interfaces over time.

\end{document}